\documentclass[twocolumn,showpacs,preprintnumbers,amsmath,amssymb]{revtex4}
\usepackage{graphicx}% Include figure files
\usepackage{dcolumn}% Align table columns on decimal point
\usepackage{bm}% bold math
\usepackage{epsfig}

\newcommand{\as}{a\!\!\!/}
\newcommand{\As}{A\!\!\!/}
\newcommand{\ks}{k\!\!\!/}
\newcommand{\ps}{p\!\!\!/}
\newcommand{\ssl}{s\!\!\!/}

\begin{document}

\preprint{}

\title{Mott scattering of polarized electrons in a strong laser field}% Force line breaks with \\
%-------------------------------------------------------------------------------------

%\author{B. Manaut and S. Taj }\thanks{e-mail: attaourti@ucam.ac.ma}
%% \homepage{http://www.Second.institution.edu/~Charlie.Author}
%\affiliation{
% UFR de Physique Atomique, Mol\'eculaire et Optique
%Appliqu\'ee,
%\\Facult\'e des Sciences, Universit\'e Moulay  Isma\"{\i}l   Bo\^{\i}te
%Postale : 4010, Beni M'hamed, Mekn\`es, Maroc }
%
%\author{Y. Attaourti}
%\affiliation{ Laboratoire de Physique des Hautes Energies et
%d'Astrophysique\\
%Facult\'e des Sciences Semlalia, Universit\'e Cadi Ayyad,
%Marrakech, Bo\^{\i}te Postale : 2390, Maroc.%\textbackslash\textbackslash
%}
\author{B. Manaut}
 \altaffiliation[Electronic address: ]{bouzid.manaut@laposte.net}%Lines break automatically or can be forced with \\
\author{S. Taj}%
 %\email{Souad_taj@yahoo.fr}
\affiliation{%
UFR de Physique Atomique, Mol\'eculaire et Optique Appliqu\'ee,
\\Facult\'e des Sciences, Universit\'e Moulay  Isma\"{\i}l   Bo\^{\i}te
Postale : 4010, Beni M'hamed, Mekn\`es, Maroc
}%

\author{Y. Attaourti}
\email{attaourti@ucam.ac.ma}
 %\homepage{http://www.Second.institution.edu/~Charlie.Author}
\affiliation{ Laboratoire de Physique des Hautes Energies et
d'Astrophysique\\
Facult\'e des Sciences Semlalia, Universit\'e Cadi Ayyad,
Marrakech, Bo\^{\i}te Postale : 2390, Maroc.
}%
\begin{abstract}
We present analytical and numerical results of the relativistic
calculation of the transition matrix element $S_{fi}$ and
differential cross section for Mott scattering of initially
polarized Dirac particles (electrons) in the presence of strong
laser field with linear polarization. We use exact Dirac-Volkov
wave functions to describe the dressed electrons and the
collision process is treated in the first Born approximation. The
influence of the laser field on the degree of polarization of the
scattered electron is reported.
\end{abstract}
% PACS, the Physics and Astronomy
                             % Classification Scheme.
%\keywords{Suggested keywords}%Use showkeys class option if keyword
                              %display desired
\maketitle
%---------------------------------
\section{Introduction}
%----------------------------------
Spin is an essential and fascinating complication in the physics
of quantum collision theory as well as in many fields of physics.
The spin of a particle is a quantum mechanical attribute.
Therefore, questions about the spin dependence of atomic reactions
tend to probe the underlying theoretical structure very deeply. On
the particle side, the technology of spin measurement has improved
dramatically over the past years [1-2]. Improvement in polarized
sources allow to produce successfully polarized gas whereas
polarized electrons and positrons in ($e^{+}e^{-}$) colliders are
commonplace. This area of study has served as a crucial testing
grounds in elementary particle physics and in atomic scattering
theories and experiments. Methods that have passed this test
successfully are now being widely used for a series of practical
applications as well. These include the production of atomic data
for the modelling of fusion plasma, as well as data needed for
astrophysics and laser physics. In this contribution, we begin by
a simple introduction to the process of Mott scattering, one the
most important techniques in polarized electron studies [3].
Whenever the spin direction plays a role, one has to average over
all spin orientations in order to describe the experiments
properly. Only in recent years has it been found possible to
produce electron beams in which the spin has a preferential
orientation. They are called polarized electrons beams [4] in
analogy to polarized light in which it is the field vectors that
have a preferred orientation. There are many reasons for the
interest in polarized electrons. One important reason is that in
physical experiments one endeavors to define as exactly as
possible the initial and/or final states of the systems being
considered. Moreover experiments on laser-induced process in
strong laser fields well beyond the atomic field strength
intensity of about $3.5$ $10^{16}$ $W/cm^{2}$ have clearly given
evidence of relativistic effects [5]. However, the search for
spin-specific effects has been rare [6]-[7].

The purpose of this contribution is to add some new physical
insights and to show that the modification of the polarization
degree due to the presence of a strong laser field can provide (or
not) a clear signature of spin effects in electron-laser
interaction. Before we present the results of our investigation
concerning laser-assisted Mott scattering of polarized electrons,
we first begin by sketching the principal steps of our treatment.
As many atomic and laser physicists are unfamiliar with the
relativistic formalism that is very often given in natural units
($\hbar =c=1$) widely used in elementary particle physics, we
begin by the most basic results of Mott scattering of polarized
electrons in the absence of the laser field using atomic units
($a.u$). In atomic units, one has ($\hbar =m_{e}=e=1$) where
$m_{e}$ is the electron rest mass. Throughout this work, we shall
use
atomic units and work with the metric tensor $g^{\mu \nu }=diag(1,-1,-1,-1)$%
. Then, in the presence of a laser field, we compare the results
obtained with those obtained in the absence of the laser field.
The organization of this paper is as follows. In section II, we
present the scattering of polarized electrons by a Coulomb field
in the absence of the laser field and we introduce the concept of
polarized differential cross section (polarized DCS) as well as
the helicity flip polarized DCS and the helicity non-flip DCS. We
also define the degree of polarization of the scattered electrons.
At this stage, it is important to remark that in many experiments
[8], the degree of polarization of the scattered electrons is
measured. In section III, we discuss the laser-assisted Mott
scattering of polarized electrons in the presence of a laser
field. In section IV, we discuss the results we have obtained and
we end by a brief conclusion in section V. We hope to offer a
simple pedagogical treatment of spin in relativistic atomic
collisions that strips it of its unnecessary mystery. Our approach
based upon the helicity formalism leads to a unified treatment
that can be applied to other relativistic atomic scattering
processes in the absence or in the presence of a laser field.
Finally, we would like to emphasize the following : the spin
polarization state of a Dirac particle has not to be confused with
the polarization of the laser field (which can be linear, circular
or elliptical) used to describe the process let aside with the
number $n$ of photons exchanged between the Dirac particles and
the laser field.
\section{Mott scattering of polarized electrons in the absence of a laser
field.}

Let us consider one of the simplest process of QED [9] namely the
Coulomb scattering of electrons. We calculate the scattering of
electrons at a fixed Coulomb potential in the first Born
approximation. The transition matrix element for this process is

\begin{equation}
S_{fi}=-i\int_{-\infty }^{+\infty }dt<\psi _{p_{f}}(x)|\As| \psi
_{pi}(x)>\label{1}
\end{equation}
where
\begin{equation}
\psi
_{p_i}(x)=\frac{1}{\sqrt{2E_{i}V}}u(p_{i},s_{i})e^{-ip_{i}x}\label{2}
\end{equation}
is the Dirac wave function describing the incident electron
normalized to the volume $V$, and where
\begin{equation}
\psi
_{p_{f}}(x)=\frac{1}{\sqrt{2E_{f}V}}u(p_{f},s_{f})e^{-ip_{f}x}\label{3}
\end{equation}
describes the scattered electron and is also normalized to the
volume $V$. The free spinors $u(p,s)$ are normalized according to
$\overline{u}u=2c^{2}$ and $u^{\dagger }u=2E$. The Coulomb
potential $A_{0}(x)$ is generated by a static nucleus of charge
($-Z$)
\begin{equation}
A_{coul}^{\mu }=(-\frac{Z}{|\mathbf{x}|},0,0,0)\label{4}
\end{equation}
Then
\begin{equation}
\As=A_{\mu }\gamma ^{\mu }=A_{0}\gamma
^{0}=-\frac{Z}{|\mathbf{x}|} \gamma ^{0}\label{5}
\end{equation}
and $S_{fi}$ simply reads
\begin{equation}
S_{fi}=iZ\int_{-\infty }^{+\infty }dt\int
d\mathbf{x}\frac{\overline{u}
\mathbf{(}p_{f},s_{f}\mathbf{)}}{\sqrt{2E_{f}V}}\gamma ^{0}\frac{
u(p_{i},s_{i}\mathbf{)}}{\sqrt{2E_{i}V}}\frac{e^{-i(p_{i}-p_{f})x}}{\left|
\mathbf{x}\right| }\label{6}
\end{equation}
Performing the integrals over $t$ and $\mathbf{x}$ gives
\begin{equation}
S_{fi}=\frac{iZ4\pi}{2V\sqrt{E_{f}E_{i}}}\frac{\overline{u}(p_{f},s_{f}
)\gamma^{0}u(p_{i},s_{i})}{|\mathbf{q}|^{2}} 2\pi \delta
(E_{f}-E_{i}),\label{7}
\end{equation}
where $|\mathbf{q}|=|\mathbf{p}_i-\mathbf{p}_f|$.\\
Using the standard procedures of QED [9], one finds for the
unpolarized DCS
\begin{equation}
\frac{d\overline{\sigma }}{d\Omega
_{f}}=\frac{Z^{2}}{c^{4}}\frac{1}{|
\mathbf{q}|^{4}}\frac{1}{2}\sum_{s_{i}s_{f}}|\overline{u}\mathbf{(}
p_{f},s_{f}\mathbf{)}\gamma
^{0}u(p_{i},s_{i}\mathbf{)}|^{2}\label{8}
\end{equation}
evaluated for $E_{f}=E_{i}=E$. This in turn implies
$|\mathbf{p}_{i}|=| \mathbf{p}_{f}|=|\mathbf{p}|.$ The unpolarized
DCS is then given by summing the DCS over the final spin
polarizations $s_{f}$ and then averaging over the initial spin
polarizations $s_{i}.$ It is straightforward to show that
\begin{eqnarray}
&&\frac{1}{2}\sum_{s_{i}s_{f}}\left|
\overline{u}(p_{f},s_{f})\gamma
^{0}u(p_{i},s_{i})\right| ^{2}=2c^{2}\nonumber\\
&&\quad\quad\quad\times\left( \frac{2E_{i}E_{f}}{c^{2}}
-(p_{i}.p_{f})+c^{2}\right)\label{9}
\end{eqnarray}
or with the notational conventions we have adopted
\begin{equation}
\frac{1}{2}\sum_{s_{i}s_{f}}\left| \overline{u}(p_{f},s_{f})\gamma
^{0}u(p_{i},s_{i})\right| ^{2}=2E^{2}+2c^2\left| \mathbf{p}\right|
^{2}\cos (\theta )+2c^{4}\label{10}
\end{equation}
where $\theta $ is the scattering angle. Using $\beta
^{2}E^{2}=\left| \mathbf{p}\right| ^{2}c^{2}$ and $\mathbf{q}$
=$\left| \mathbf{p}_{i}- \mathbf{p}_{f}\right| =2\left|
\mathbf{p}\right| \sin (\theta /2)$, one finally finds for the
unpolarized DCS
\begin{eqnarray}
\left( \frac{d\overline{\sigma }}{d\Omega _{f}}\right) _{Mott}
&=&\frac{1}{4} \frac{Z^{2}}{c^{2}\beta ^{2}\left|
\mathbf{p}\right| ^{2}}\frac{(1-\beta ^{2}\sin ^{2}(\frac{\theta
}{2}))}{\sin ^{4}(\frac{\theta }{2})} \\ \label{11}
 &=&\left(
\frac{d\overline{\sigma }}{d\Omega _{f}}\right) _{Ruth}\left(
1-\beta ^{2}\sin ^{2}(\frac{\theta }{2})\right) \nonumber
\end{eqnarray}
where
\begin{equation}
\left( \frac{d\overline{\sigma }}{d\Omega _{f}}\right)
_{Ruth}=\frac{Z^{2}}{ 4c^{2}\beta ^{2}\left| \mathbf{p}\right|
^{2}}\frac{1}{\sin ^{4}(\frac{ \theta }{2})}\label{12}
\end{equation}
is the unpolarized Rutherford DCS obtained in the limit $\beta
\rightarrow 0$ (small velocities). Up to now, all calculations
have been carried out with the assumption that the electron spin
is not observed. In what follows, we shall calculate the
scattering of polarized Dirac particles. But first, let us review
the basics needed for the description of spin polarization [9].
Free electrons with four-momentum $p$ and spin $s$ are described
by the free spinors $u(p,s)$, the vector $s^{\mu }$ defined by
\begin{equation}
s^{\mu }=\frac{1}{c}\left( \left| \mathbf{p}\right| ,\frac{E}{c}\widehat{%
\mathbf{p}}\right),\label{13}
\end{equation}
(with $\widehat{\mathbf{p}}=\mathbf{p}/|\mathbf{p}|$ ) is a
Lorentz vector in a frame in which the particle moves with
momentum $\mathbf{p}$. One easily checks the normalization
condition
\begin{equation}
s.s=s_{\mu }.s^{\mu }=-1,\label{14}
\end{equation}
and the orthogonality condition
\begin{equation}
p.s=p_{\mu }.s^{\mu }=0.\label{15}
\end{equation}
We introduce the spin projection operator
\begin{equation}
\widehat{\Sigma }(s)=\frac{1}{2}(1+\gamma _{5}\ssl),\label{16}
\end{equation}
with $\gamma _{5}=i\gamma ^{0}\gamma ^{1}\gamma ^{2}\gamma
^{3}=-i\gamma _{0}\gamma _{1}\gamma _{2}\gamma _{3}$. This
operator has the properties that
\begin{equation}
\widehat{\Sigma }(s)u(p,\pm s)=\pm  u(p,s).\label{17}
\end{equation}
One can also apply this formalism to helicity states where the
spin points in the direction of the momentum $\mathbf{p}$
\begin{equation}
s_{\lambda }^{\prime}=\lambda \frac{\mathbf{p}}{\left|
\mathbf{p}\right| }=\lambda \text{ }\widehat{\mathbf{p}},\text{
}\lambda =\pm 1.\label{18}
\end{equation}
We can then define a four spin vector
\begin{equation}
s_{\lambda }^{\mu }=\frac{\lambda }{c}\left( \left|
\mathbf{p}\right| ,\frac{
E}{c}\widehat{\mathbf{p}}\right).\label{19}
\end{equation}
The starting point of our calculation is the DCS for Coulomb
scattering of an electron with well defined momentum $p_{i}$ and
well defined spin $s_{i}$. If also the final spin $s_{f}$ is
measured, the polarized DCS then reads
\begin{equation}
\frac{d\sigma }{d\Omega _{f}}=\frac{Z^{2}}{c^{4}}\frac{\left|
\overline{u}( p_{f},s_{f})\gamma ^{0}u(p_{i},s_{i})\right|
^{2}}{\left| \mathbf{q} \right| ^{4}},\label{20}
\end{equation}
evaluated for $E_{f}=$ $E_{i}=E$. Introducing the operators
\begin{equation}
\widehat{\Sigma }_{\lambda _{i}}(s)=\frac{1}{2}(1+\lambda
_{i}\gamma _{5}\ssl_{i}),\label{21}
\end{equation}

\begin{equation}
\widehat{\Sigma }_{\lambda _{f}}(s)=\frac{1}{2}(1+\lambda
_{_{f}}\gamma _{5}\ssl_{f}),\label{22}
\end{equation}
we obtain for the polarized DCS
\begin{widetext}
\begin{eqnarray}
\frac{d\sigma }{d\Omega _{f}}(\lambda _{i},\lambda
_{_{f}})=\frac{Z^{2}}{ c^{4}\left| \mathbf{q}\right|
^{4}}\text{Tr} \Big\{ \gamma ^{0}\frac{(1+\lambda _{i}\gamma
_{5}\ssl_{i})}{2} (\ps_{i}c+c^{2})\gamma ^{0}\frac{(1+\lambda
_f\gamma _{5}\ssl_{f})}{2}(\ps_{f}c+c^{2})\Big\}\label{23}
\end{eqnarray}
\end{widetext}
 Using the relations
$p_{i}s_{f}=p_{f}s_{i}=E\left| \mathbf{p}\right| (1-\cos (\theta
))/c^{2}$, $s_{i}s_{f}=(\left| \mathbf{p}\right| ^{2}-E^{2}\cos
(\theta )/c^{2})/c^{2}$ and $p_{i}p_{f}=E^{2} /c^{2}-\left|
\mathbf{p}\right| ^{2}\cos (\theta )$, one finally obtains
\begin{eqnarray}
&&\frac{d\sigma }{d\Omega _{f}}(\lambda _{i},\lambda
_{_{f}})=\frac{Z^{2}}{ c^{4}\left| \mathbf{q}\right| ^{4}}2\Big[
E^{2}\cos ^{2}(\frac{\theta }{2} )+c^{4}\sin ^{2}(\frac{\theta
}{4})\nonumber\\
&&\quad\quad+\lambda _{i}\lambda _{f}\left(E^{2}\cos
^{2}(\frac{\theta }{2})-c^{4}\sin ^{2}(\frac{\theta
}{4})\right)\Big].\label{24}
\end{eqnarray}
So if during the process, there is a helicity flip $\lambda
_{f}=-\lambda _{i}$, we obtain for the helicity flip polarized DCS
\begin{eqnarray}
\left( \frac{d\sigma }{d\Omega _{f}}\right) _{flip}
&=&\frac{4Z^{2}}{ c^{4}\left| \mathbf{q}\right| ^{4}}c^{4}\sin
^{2}(\frac{\theta }{2}) \\ \label{25} &=&\left(
\frac{d\overline{\sigma} }{d\Omega _{f}}\right)
_{Ruth}\frac{c^{4}}{E^{2}} \sin ^{2}(\frac{\theta }{2}). \nonumber
\end{eqnarray}
If there is no helicity flip $\lambda _{i}=\lambda _{f}$ and the
helicity non flip polarized DCS is given
\begin{equation}
\left( \frac{d\sigma }{d\Omega _{f}}\right) _{non\text{
}flip}=\left( \frac{ d\overline{\sigma} }{d\Omega _{f}}\right)
_{Ruth}\cos ^{2}(\frac{\theta }{2}),\label{26}
\end{equation}
Of course, one must have
\begin{equation}
\left( \frac{d\overline{\sigma }}{d\Omega _{f}}\right)
_{Mott}=\left( \frac{ d\sigma }{d\Omega _{f}}\right) _{non\text{
}flip}+\left( \frac{d\sigma }{ d\Omega _{f}}\right) _{\text{
}flip}.\label{27}
\end{equation}
Noting that $c^{4}=\left| \mathbf{p}\right| ^{2}c^{2}-E^{2}$, we
have
\begin{eqnarray}
\frac{c^{4}}{E^{2}}\sin ^{2}(\frac{\theta }{4})+\cos
^{2}(\frac{\theta }{2}) &=&1-\frac{\left| \mathbf{p}\right|
^{2}c^{2}}{E^{2}}\sin ^{2}(\frac{\theta }{2}) \\
\label{28} &=&1-\beta ^{2}\sin ^{2}(\frac{\theta }{2}),  \nonumber
\end{eqnarray}
and we end up with
\begin{equation}
\left( \frac{d\overline{\sigma }}{d\Omega _{f}}\right)
_{Mott}=\left( \frac{d \overline{\sigma }}{d\Omega _{f}}\right)
_{Ruth}\left( 1-\beta ^{2}\sin ^{2}( \frac{\theta
}{2})\right),\label{29}
\end{equation}
Now, we introduce the degree of polarization $P$ which is defined
by
\begin{equation}
P =\frac{\frac{d\sigma }{d\Omega _{f}}(\lambda _{_{f}}=\lambda
_{i}=1)- \frac{d\sigma }{d\Omega _{f}}(\lambda _{_{f}}=-\lambda
_{i}=1)}{\frac{ d\sigma }{d\Omega _{f}}(\lambda _{_{f}}=\lambda
_{i}=1)+\frac{d\sigma }{ d\Omega _{f}}(\lambda _{_{f}}=-\lambda
_{i}=1)}\label{30}
\end{equation}
For our process, this degree of polarization reads
\begin{equation}
P=1-\frac{c^{4}\sin ^{2}(\frac{\theta }{2})}{E^{2}\cos
^{2}(\frac{\theta }{2})+c^{4}\sin ^{2}(\frac{\theta
}{2})}\label{31}
\end{equation}
Remarking that $E=\gamma c^{2}$ where $\gamma $ is the
relativistic parameter $\gamma =(1-\beta ^{2})^{-\frac{1}{2}}$, we
obtain for $P$
\begin{equation}
P =1-\frac{2\sin ^{2}(\frac{\theta }{2})}{\gamma ^{2}\cos
^{2}(\frac{ \theta }{2})+\sin ^{2}(\frac{\theta }{2})}\label{32}
\end{equation}
So, for the process of Mott scattering in the absence of the laser
field, only two parameters are relevant, the relativistic
parameter $\gamma $ and the scattering angle $\theta$. To end this
section, let us remark that the kinetic energy of the incident
electron is given by
\begin{equation}
T_{i}=T_{f}=c^{2}(\gamma -1)\label{33}
\end{equation}
\section{Mott scattering of polarized electrons in the presence of a
strong laser field with linear polarization.}
 Let us consider the
simple case where the linearly polarized laser field is described
by the four potential
\begin{equation}
A(x)=a_{1}\cos (k.x)\label{34}
\end{equation}
with $k.x=k_{\mu }x^{\mu }=\omega t-\mathbf{k.x}$ where $\omega $
and $ \mathbf{k}$ are the frequency and the wave vector of the
laser field respectively. The electron is described by the Dirac
-Volkov wave function [10] solution of the Dirac equation in the
presence of a laser field having the following form
\begin{eqnarray}
\psi _{p}(x) &=&\psi _{q}(x)=\left( 1+\frac{\ks\As}{
2c(k.p)}\right) \frac{u(p,s)}{\sqrt{2QV}}\exp (iS(x))\nonumber \\
&=&\left( 1+\frac{\ks\as_1}{2c(k.p)}\cos (\phi )\right)
\frac{u(p,s)}{\sqrt{2QV}}\exp (iS(x)),\label{35}
\end{eqnarray}
with
\begin{equation}
S(x)=-q.x-\frac{a_{1}.p}{c(k.p)}\sin (\phi )\label{36}
\end{equation}
\begin{equation}
q^{\mu }=p^{\mu }-\frac{\overline{A}^{2}}{2c^{2}(k.p)}k^{\mu
}\label{37}
\end{equation}
where $\overline{A}^{2}=\overline{a_{1}^{2}}=-\mathbf{a}^{2}/2$ is
the averaged squared potential. We turn now to the calculation of
the transition amplitude. The interaction of the dressed electron
with the central Coulomb field
\begin{equation}
A_{Coul}^{\mu }=\left( -\frac{Z}{\left| \mathbf{x}\right| },0,0,0
\right)\label{38}
\end{equation}
is considered as a first order perturbation. This is well
justified if  $Z\alpha \ll 1$ where $Z$ is the atomic number and
$\alpha $ is the fine structure constant. Then, the transition
matrix element for the transition $(i\rightarrow f)$ is
\begin{equation}
S_{fi}=iZ\int_{-\infty }^{+\infty }dt\langle \psi
_{q_{_{f}}}(x)\frac{\gamma ^{0}}{\left| \mathbf{x}\right| }\psi
_{q_{_{i}}}(x)\rangle \label{39}
\end{equation}
Let us consider the term
\begin{eqnarray}
\overline{\psi }_{q_{_{f}}}(x)\gamma ^{0}\psi
_{q_{_{i}}}(x)&=&\frac{\overline{
u}(p_{f},s_{f})}{\sqrt{2Q_{f}V}}\left[ 1+c(p_{f})\as_{1}\ks\cos
(\phi )\right]\nonumber\\
&\times& \gamma ^{0}\left[ 1+c(p_{i})\ks\as_{1}\cos (\phi
)\right]\frac{u(p_{i},s_{i})}{\sqrt{2Q_{i}V}} \nonumber\\
&\times&\exp (-i(q_{i}-q_{f}).x-iz\sin (\phi ))\label{40}
\end{eqnarray}
with
\begin{equation}
c(p)=\frac{1}{2c(k.p)}\quad;\quad z=\frac{1}{c}\left(
\frac{a_{1}.p_{i}}{k.p_{i}}-\frac{a_{1}.p_{f}}{k.p_{f}}
\right)\label{41}
\end{equation}
Transforming the terms containing products of Dirac gamma matrices
and invoking the well known result for ordinary Bessel functions
\begin{equation}
e^{[-iz\sin (\phi )]}=\sum_{n=-\infty }^{+\infty
}J_n(z)e^{(-in\phi )},
\end{equation}
one ends up with
\begin{eqnarray}
&&S_{fi}=\frac{iZ4\pi}{\sqrt{ 2Q_{i}V}\sqrt{
2Q_{f}V}}\sum_{n=-\infty }^{+\infty
}\frac{\overline{u}(p_{f},s_{f})\Gamma _{n}u(p_{i},s_{i})}{\left|
\mathbf{q}_{i}+n\mathbf{k-q}_{f}\right| ^{2}}\nonumber\\
&&\quad\quad \times 2\pi \delta (Q_{i}-Q_{f}+n\omega )\label{42}
\end{eqnarray}
with
\begin{equation}
\Gamma _{n}=\gamma ^{0}B_{0n}+\gamma
^{0}\ks\as_1B_{1n}+\as_{1}\ks\gamma ^{0}B_{2n}-2k^{2}a^{2}\ks
B_{3n}\label{43}
\end{equation}
where  $B_{0n},$ $B_{1n},$ $B_{2n}$  and $B_{3n}$ are given by
\begin{equation}
\left\{\begin{array}{l}
B_{0n}=J_{n}(z) \\
\\
B_{1n}=\frac{nJ_{n}(z)}{2c(k.p_{i})z}=\frac{1}{4c(k.p_{i})}\left[
J_{n+1}(z)+J_{n-1}(z)\right] \\
\\
B_{2n}=\frac{nJ_{n}(z)}{2c(k.p_{f})z}=\frac{1}{4c(k.p_{f})}\left[
J_{n+1}(z)+J_{n-1}(z)\right] \\
\\
B_{3n}=\frac{J_{n-2}(z)+2J_{n}(z)+J_{n+2}(z)}{16c^2(k.p_{i})(k.p_{f})}
\end{array}\right\}\label{44}
\end{equation}
Using the standard procedures of QED, one gets for the unpolarized
DCS
\begin{equation}
\frac{d\overline{\sigma} }{d\Omega _{f}}=\sum_{n=-\infty
}^{+\infty }\frac{d \overline{\sigma}^{n}}{d\Omega _{f}}\label{45}
\end{equation}
with
\begin{equation}
\frac{d\overline{\sigma}^{n}}{d\Omega
_{f}}=\frac{Z^{2}}{c^{4}}\frac{\left| \mathbf{q}_{f}\right|
}{\left| \mathbf{q}_{i}\right| }\frac{1}{\left|
\mathbf{q}_{i}+n\mathbf{k}-\mathbf{q}_{f}\right| ^{4}}\frac{1}{2}\sum_{s_{i}s_{f}}%
\left| M_{fi}^{(n)}\right|^{2}_{Q_{f}=Q_{i}+n\omega }\label{46}
\end{equation}
where
%\begin{widetext}
\begin{equation}
\frac{1}{2}\sum_{s_{i}s_{f}}\left| M_{fi}^{(n)}\right|
^{2}=\frac{1}{2} \text{Tr}\left\{
\Gamma_{n}(c\ps_{i}+c^{2})\overline{\Gamma
}_{n}(c\ps_{f}+c^{2})\right\}\label{47}
\end{equation}
%\end{widetext}
whereas the corresponding polarized DCS is obtained by introducing
the operators given in Eqs. (\ref{21}),(\ref{22})
\begin{equation}
\frac{d\sigma }{d\Omega_{f}}(\lambda _{i},\lambda
_{_{f}})=\sum_{n=-\infty }^{+\infty }\frac{d\sigma^{n}}{d\Omega
_{f}}(\lambda _{i},\lambda_{f})\label{48}
\end{equation}
with
\begin{widetext}
\begin{equation}
\frac{d\sigma^{n}}{d\Omega _{f}}(\lambda _{i},\lambda
_{_{f}})=\frac{Z^{2}}{ c^{4}}\frac{\left| \mathbf{q}_{f}\right|
}{\left| \mathbf{q}_{i}\right| } \frac{1}{\left|
\mathbf{q}_{i}+n\mathbf{k}-\mathbf{q}_{f}\right|
^{4}}\text{Tr}\left\{ \Gamma _{n}\frac{(1+\lambda _{i}\gamma
_{5}\ssl_{i})}{2}(c\ps_{i}+c^{2})\overline{\Gamma }_{n}\frac{
(1+\lambda _{f}\gamma
_{5}\ssl_{f})}{2}(c\ps_{f}+c^{2})\right\}\label{49}
\end{equation}
\end{widetext}
In equations (\ref{47}) and (\ref{49})
\begin{equation}
\overline{\Gamma }_{n}=\gamma ^{0}\Gamma _{n}^{\dagger }\gamma
^{0}\label{50}
\end{equation}
Using REDUCE [11], one finds that the polarized DCS
\begin{equation}
\frac{d\sigma }{d\Omega _{f}}(\lambda _{i},\lambda
_{_{f}})=\sum_{n=-\infty}^{+\infty}\frac{Z^2|\mathbf{q}_f|}{c^4|\mathbf{q}_i|}\frac{(\mathcal{A}_n+\lambda
_{i}\lambda
_{_{f}}\mathcal{B}_n)}{\left|\mathbf{q}_{i}+n\mathbf{k}-\mathbf{q}_{f}\right|
^{4}}\label{51}
\end{equation}
where $\mathcal{A}_n$ and $\mathcal{B}_n$ are given by
\begin{eqnarray}
\mathcal{A}_n&=&A_1B^2_{0n}+B_1B^2_{1n}+C_1B^2_{2n}+D_1B^2_{3n}+E_1B_{0n}B_{1n}\nonumber\\
&&+F_1B_{0n}B_{2n}+G_1B_{0n}B_{3n}+H_1B_{1n}B_{2n}\nonumber\\
&&+X_1B_{1n}B_{3n}+Y_1B_{2n}B_{3n},
\end{eqnarray}

\begin{eqnarray}
\mathcal{B}_n&=&A_2B^2_{0n}+B_2B^2_{1n}+C_2B^2_{2n}+D_2B^2_{3n}+E_2B_{0n}B_{1n}\nonumber\\
&&+F_2B_{0n}B_{2n}+G_2B_{0n}B_{3n}+H_2B_{1n}B_{2n}\nonumber\\
&&+X_2B_{1n}B_{3n}+Y_2B_{2n}B_{3n},
\end{eqnarray}
These coefficients are very lengthy and can be found in the
Appendix. So, the helicity non flip polarized DCS is $(\lambda
_{_{f}}=-\lambda _{i}=1)$

\begin{equation}
\left( \frac{d\sigma }{d\Omega _{f}}\right)
_{flip}=\sum_{n=-\infty}^{+\infty}\frac{Z^2|\mathbf{q}_f|}{c^4|\mathbf{q}_i|}\frac{(\mathcal{A}_n-\mathcal{B}_n)}{\left|\mathbf{q}_{i}+n\mathbf{k}-\mathbf{q}_{f}\right|
^{4}} \label{52}
\end{equation}
and the helicity non flip polarized DCS is $(\lambda
_{_{f}}=\lambda _{i}=1)$

\begin{equation}
\left( \frac{d\sigma }{d\Omega _{f}}\right) _{non\text{
}flip}=\sum_{n=-\infty}^{+\infty}\frac{Z^2|\mathbf{q}_f|}{c^4|\mathbf{q}_i|}\frac{(\mathcal{A}_n+\mathcal{B}_n)}{\left|\mathbf{q}_{i}+n\mathbf{k}-\mathbf{q}_{f}\right|
^{4}}\label{53}
\end{equation}
Therefore, we obtain for the degree of polarization
\begin{equation}
P =\frac{(\frac{d\sigma }{d\Omega _{f}})_{non\text{ }flip}-
(\frac{d\sigma }{d\Omega _{f}})_{flip}}{(\frac{ d\sigma }{d\Omega
_{f}})_{non\text{ }flip}+(\frac{d\sigma }{ d\Omega
_{f}})_{flip}}\label{54}
\end{equation}
It is easy to show both analytically and numerically that the
unpolarized DCS given in (\ref{45}) is such that

\begin{equation}
\frac{d\overline{\sigma }}{d\Omega _{f}}=\left( \frac{d\sigma
}{d\Omega _{f}} \right) _{non\text{ }flip}+\left( \frac{d\sigma
}{d\Omega _{f}}\right) _{flip}\label{55}
\end{equation}
\section{Results and discussions}
\subsection{In the absence of the laser field}
We begin our discussion by one of the most fundamental result
concerning the degree of polarization in the non relativistic
regime. As it is well known, this degree is close to
$cos(\widehat{\theta _i,\theta_f})$, the scattering angle. This
degree varies from $-1$ to $1$ as the scattering angle
$\theta_{if}=(\widehat{\theta _i,\theta_f})$ varies from
$-180^{\circ}$ to $180^{\circ}$.
\begin{figure}[h]
 \begin{minipage}[b]{.46\linewidth}
  \centering\epsfig{figure=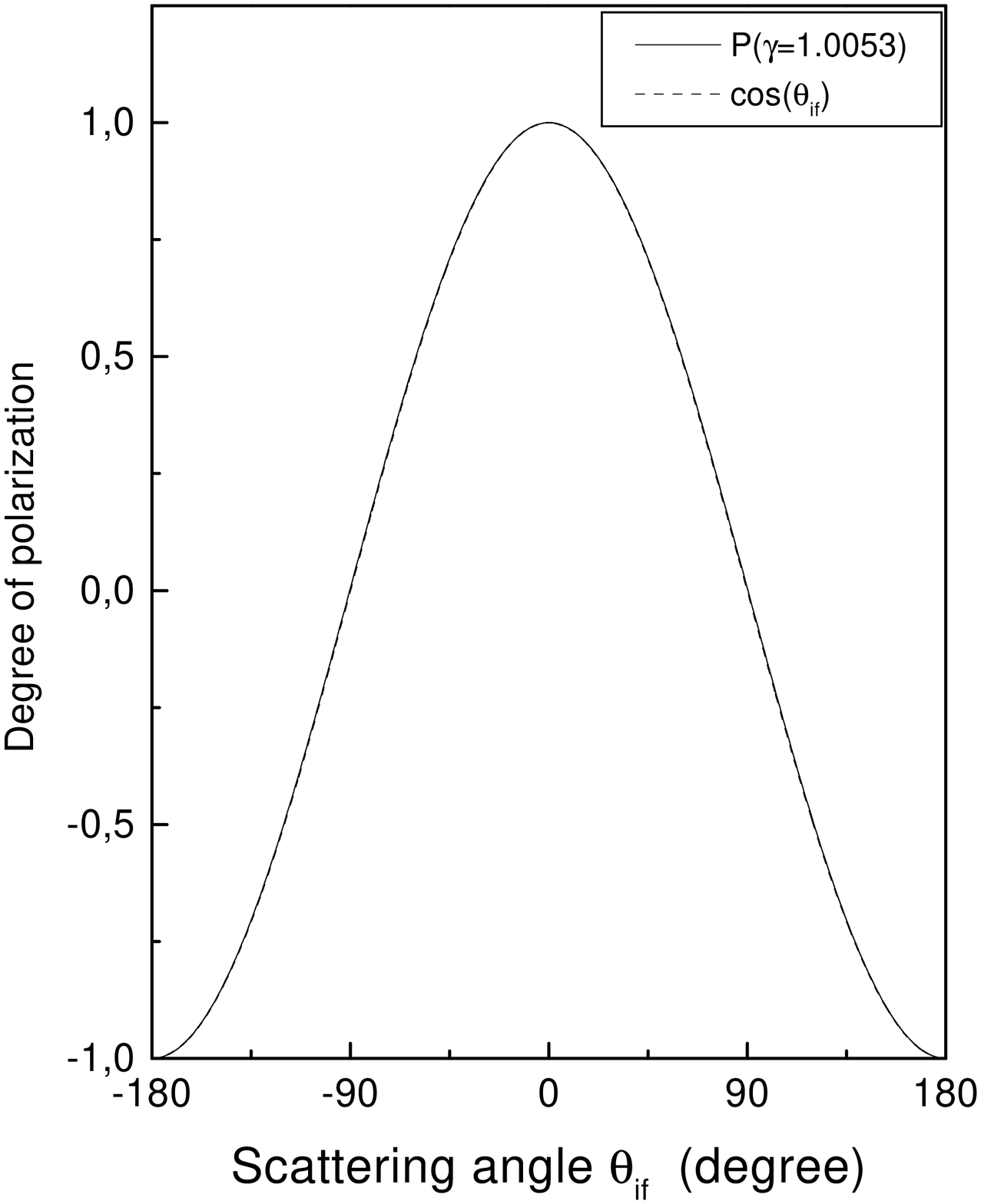,width=\linewidth,height=2.5in}
  \caption{\label{fig1} The behaviors of the degree of polarization as
  well as the cosine of the scattering angle $cos(\theta_{if})$ as functions of the
  scattering angle $\theta_{if}$. \vspace{0.64cm}}
 \end{minipage} \hfill
 \begin{minipage}[b]{.46\linewidth}
  \centering\epsfig{figure=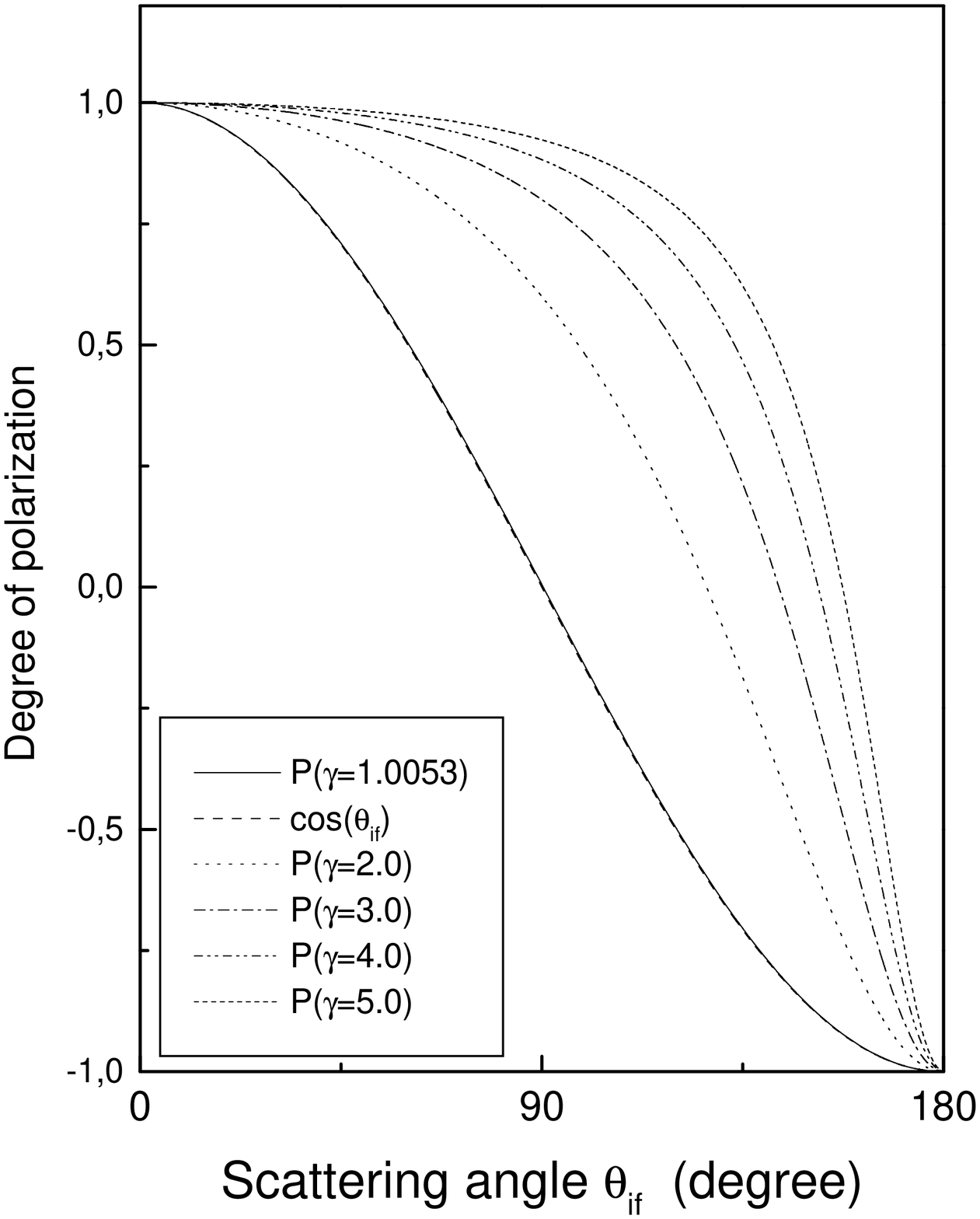,width=\linewidth,height=2.5in}
  \caption{\label{fig2}The behaviors of the degree of polarization as
  well as the cosine of the scattering angle $cos(\theta_{if})$ as functions of the
  scattering angle $\theta_{if}$, with different values of the relativistic parameter $\gamma $.}
 \end{minipage}
\end{figure}
We show in Fig. 1, the behavior of $cos(\widehat{\theta
_i,\theta_f})$ as well as that of the degree of polarization. We
obtain two very close curves, which was to be expected. With
increasing values of the relativistic parameter $\gamma$, i.e when
the collision becomes more relativistic, the degree of
polarization becomes less strongly angle dependent and approaches
a constant value $P=1$ as $E \rightarrow \infty$. This is shown in
Fig. 2. All these results are consistent with well known results
of QED [9] and to compare with results in presence of the laser
field, one has to bear in mind that the laser field dresses the
angular coordinates [12] of the incident and scattered electrons
so instead of plotting the degree of polarization as a function
of the angle $\theta_{if}$, one has to plot it as a function of
the angle $\theta_f$ of the scattered electron. The geometry we
have chosen both for the non relativistic and relativistic
regimes is the following : the angles of the incident electron
are $\theta_i=45^{\circ}$ and $\phi_i=0^{\circ}$ while the angles
of the scattered electron are $\phi_f=90^{\circ}$ with $\theta_f$
varying from $-180^{\circ}$ to $180^{\circ}$. There is still a
very good agreement between the degree of polarization $P$ and
$cos(\widehat{\theta _i,\theta_f})$ but with the difference that
now the maximum and minimum are no longer $-1$ and $1$ but nearly
$-0.7$ and $0.7$. This is shown in Fig. 3. As the energy is
increased, we observe the same phenomenon as previously
mentioned, that is, the degree of polarization approaches the same
constant value $1$ and becomes less angle dependent as this is
shown in Fig. 4.
\begin{figure}[h]
 \begin{minipage}[b]{.46\linewidth}
  \centering\epsfig{figure=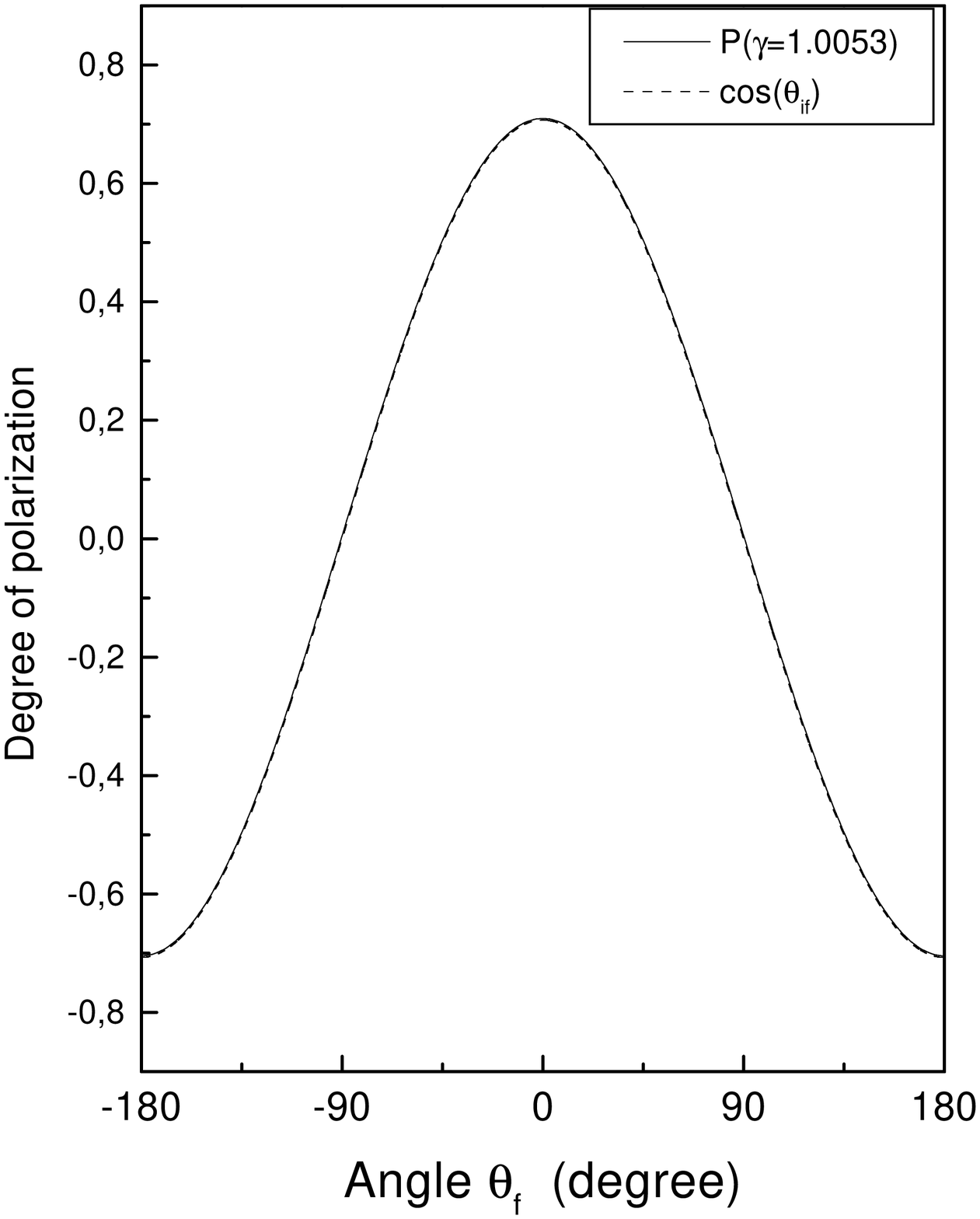,width=\linewidth,height=2.5in}
  \caption{The behaviors of the degree of polarization as
  well as the cosine of the scattering angle $cos(\theta_{if})$ as functions of the
  angle $\theta_{f}$. \vspace{0.3cm}\label{fig3}}
 \end{minipage} \hfill
 \begin{minipage}[b]{.46\linewidth}
  \centering\epsfig{figure=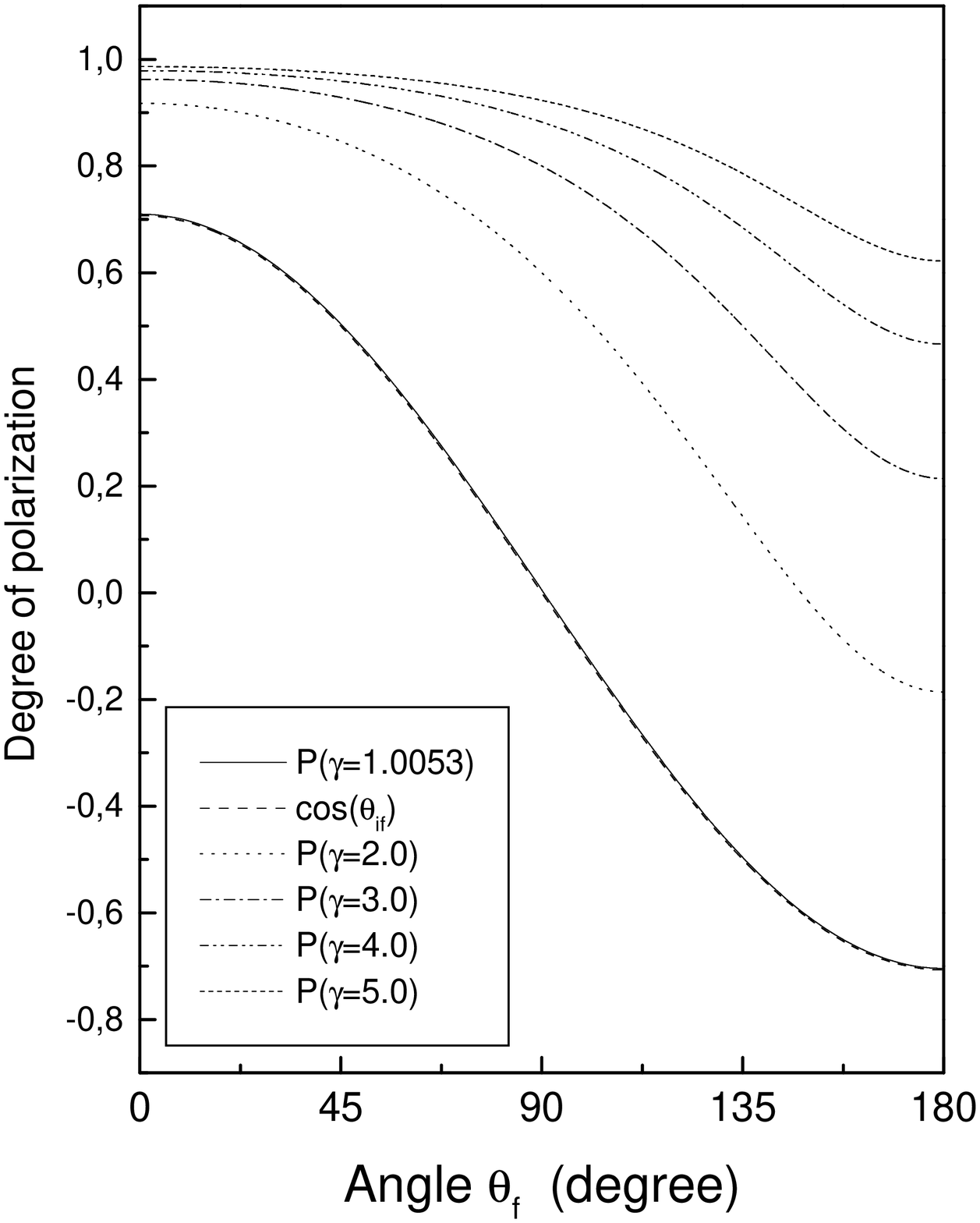,width=\linewidth,height=2.5in}
  \caption{The behaviors of the degree of polarization and $cos(\theta_{if})$ as functions of the
   angle $\theta_{f}$, with different values of the relativistic parameter $\gamma $. \label{fig4}}
 \end{minipage}
\end{figure}
One of the most striking effect we shall soon encounter in the
next section is that the laser field has a very negligible
influence on the degree of polarization while it has an influence
on the polarized DCSs. \\
Remark : in figures (2) and (4) we have plotted the degree of
polarization and $cos(\theta_{if})$ as functions of the angle
$\theta_f$ varying from $0^{\circ}$ to  $180^{\circ}$ since there
is a complete symmetry between its behavior from $0^{\circ}$ to
$180^{\circ}$ and from $-180^{\circ}$ and $0^{\circ}$.
\subsection{In the presence of the laser field}
When a laser field is present, one has to be careful to
distinguish between two regimes, the non relativistic regime and
the relativistic regime. The non relativistic regime corresponds
to a relativistic parameter $\gamma =1.0053$ where the curves for
the relativistic and non relativistic kinetic energies of a
particle begin to be distinguished. We recall that $\gamma
=(1-v^2/c^2)^{-1/2}$ where $v$ is the velocity of the particle and
$c$ is the velocity of light in the vacuum. In this regime, the
electric field strength is $E=0.05\,\,(a.u)$ and the laser
frequency is $\omega =0.043\,\, (a.u)$ which corresponds to a
laser photon energy of $\hbar \omega =1.17 \,\,eV$. The
relativistic regime corresponds to a relativistic parameter
$\gamma =2.00$. In this regime, the electric field strength is
$E=1.0\,\, (a.u)$ and we retain the same laser frequency as
previously.
\subsubsection{The non relativistic regime}
In this regime, one expects that the dressing of angular
coordinates of $\mathbf{p}_i$ and $\mathbf{q}_i$ as well as those
of $\mathbf{p}_f$ and $\mathbf{q}_f$ will not be important and
this is indeed the case. While the helicity non flip polarized
differential cross section (in short DCS $(\uparrow)$) and the
helicity flip polarized differential cross section (DCS
$(\downarrow)$) are influenced by the number of photons exchanged,
it is not the case for the degree of polarization which remains
nearly constant until the cut off is reached. We recall that when
the cut off is reached, the various DCSs do not vary anymore since
the arguments of the ordinary Bessel functions become close to
their indices [7]. This situation is shown in Fig. 5 where we
give the three DCSs : DCS $(\uparrow)$, DCS $(\downarrow)$ and the
unpolarized DCS. For every simulation and for any number of
photons exchanged, the sum of DCS $(\uparrow)$ and DCS
$(\downarrow)$ always gives the unpolarized DCS. In Fig. 5, the
number of photons exchanged is $\pm 100$ and the various DCSs are
reduced.
\begin{figure}[h]
 \begin{minipage}[b]{.46\linewidth}
  \centering\epsfig{figure=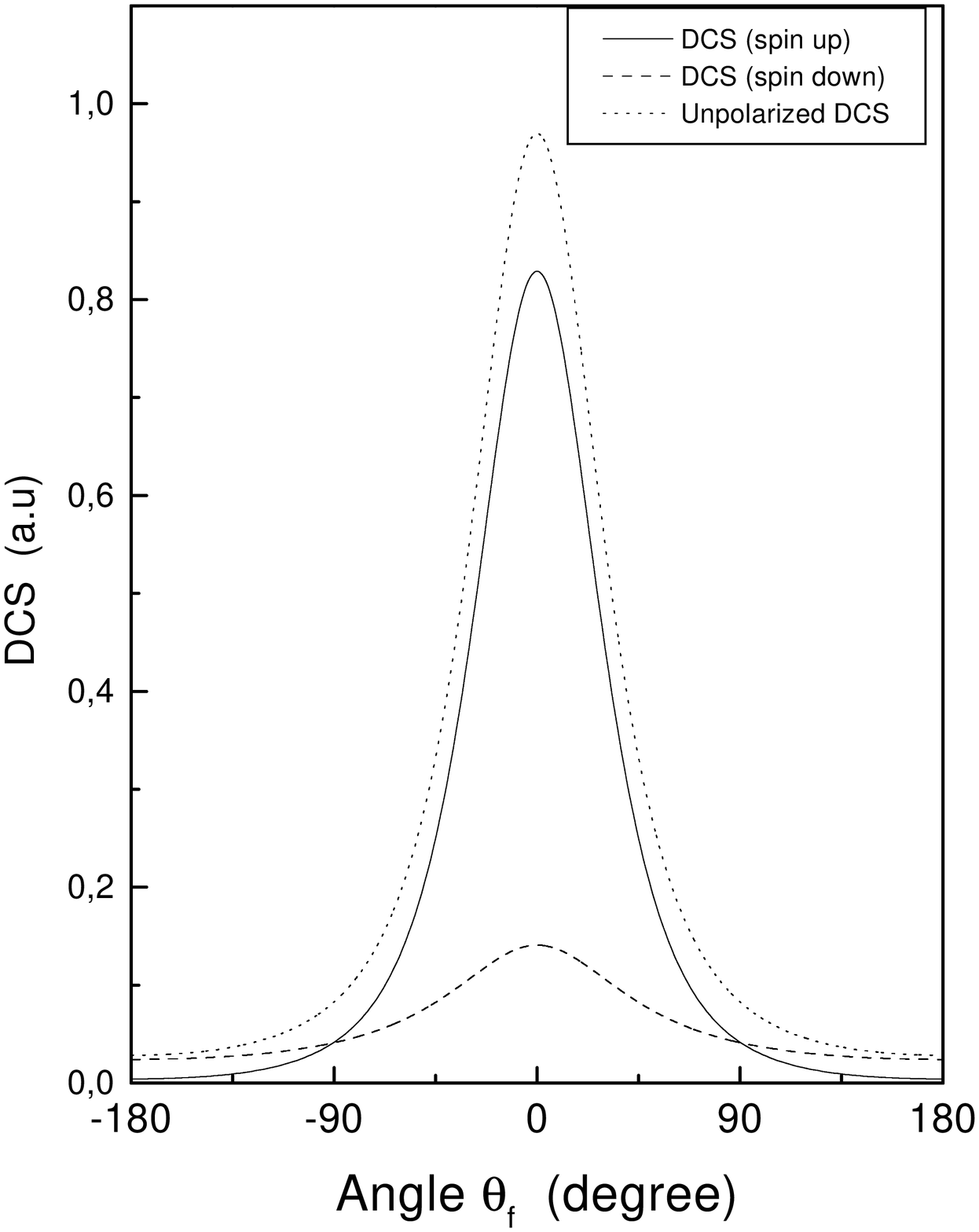,width=\linewidth,height=2.5in}
  \caption{The three DCSs : DCS $(\uparrow)$, DCS $(\downarrow)$ and unpolarized DCS scaled
   in $10^{-4}$ as functions of the angle $\theta_{f}$ in degree for  $n=\pm 100$.  \vspace{0.3cm} \label{fig5}}
 \end{minipage} \hfill
 \begin{minipage}[b]{.46\linewidth}
  \centering\epsfig{figure=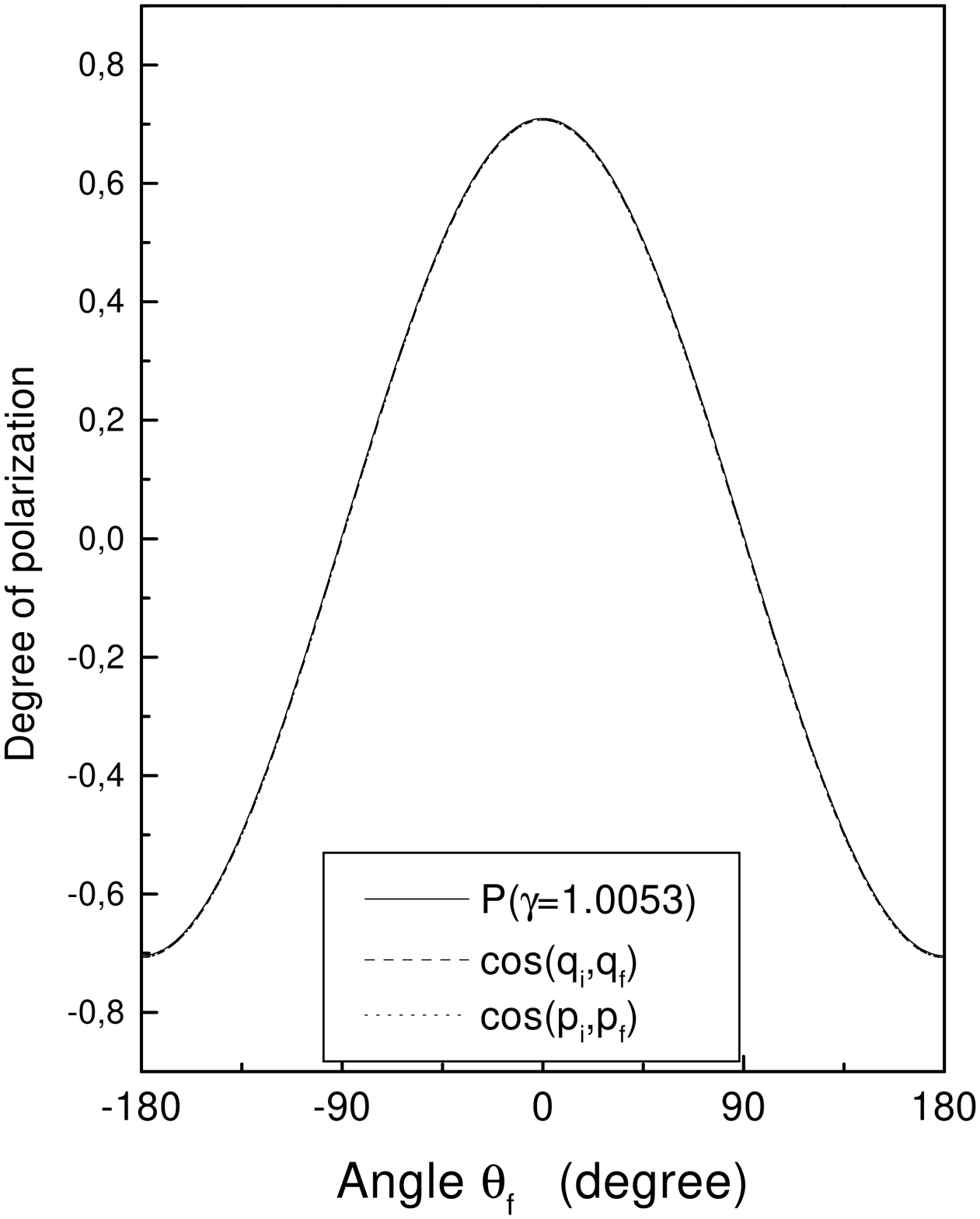,width=\linewidth,height=2.5in}
  \caption{\small{The variation of the degree of polarization $P(\gamma=1.0053)$,
   $cos(\mathbf{q}_i,\mathbf{q}_f)$ and $cos(\mathbf{p}_i,\mathbf{p}_f)$
    as functions of the angle $\theta_f$ in degree.}\hspace{2cm} \label{fig6}}
 \end{minipage}
\end{figure}
 In Fig. 6, we give the
corresponding degree of polarization and there is no difference
between this degree of polarization and that shown in Fig. 3.
When the number of photons exchanged reaches the cut off, which in
the non relativistic regime is $\pm 300$, the three DCSs increase
as shown in Fig. 7 but the degree of polarization do not increase
at all.
\begin{figure}[ht]
\centering
\includegraphics[angle=0,width=6cm,height=2.5in]{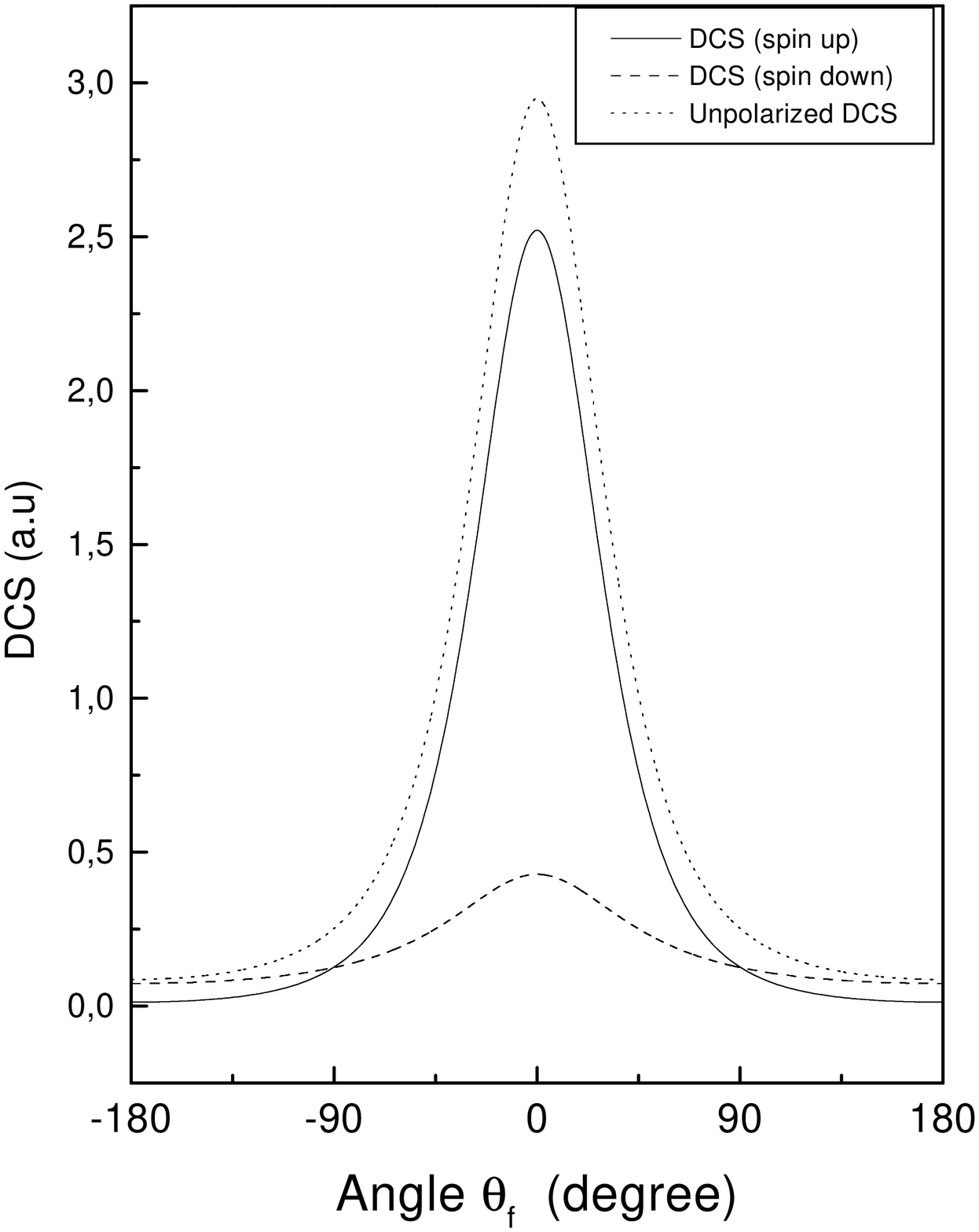}
\caption{The three DCSs : DCS $(\uparrow)$, DCS $(\downarrow)$ and
unpolarized DCS scaled in $10^{-4}$ as functions of the angle
$\theta_{f}$ in degree for an exchange of $\pm 300$ photons.   }
\end{figure}
It is still close to $cos(\widehat{\theta _i,\theta_f})$ and
increasing the number of photons has no influence on its value. If
such a behavior holds for the relativistic regime, then at least
to first order in perturbation theory, we can draw the following
conclusions : while the various DCSs are reduced by the influence
of the laser field with increasing values of the field strength as
well as the incoming electron relativistic kinetic energies, the
corresponding degrees of polarization are almost insensitive to
the presence of the laser field or in other terms to the number of
photons exchanged.

\subsubsection{The relativistic regime}
In this regime, the cut off is $\pm 83000$ photons that can be
exchanged with the laser field and the corresponding DCSs are
reduced. They are of the order $10^{-11}$ (for the laser free case
they are of the order $10^{-8}$) and we show in Fig. 8 the
behavior of the three DCS. As the energy of the incident electron
is increased (and also the electric field strength), the helicity
flip polarized differential cross section ( DCS $(\downarrow)$ is
almost negligible, showing that the leading process in this regime
is likely to favor the helicity non flip polarized differential
cross section DCS $(\uparrow)$. For an exchange of $\pm 100$
photons and for an exchange of $\pm 500$ photons, the situation
remains the same as to the relative importance of the two DCSs.
The increase in the value of the DCS $(\uparrow)$) is simply due
to the fact that we have summed over $n=\pm 500$ photons and this
can be shown in Fig. 9.
\begin{figure}[ht]
 \begin{minipage}[b]{.46\linewidth}
  \centering\epsfig{figure=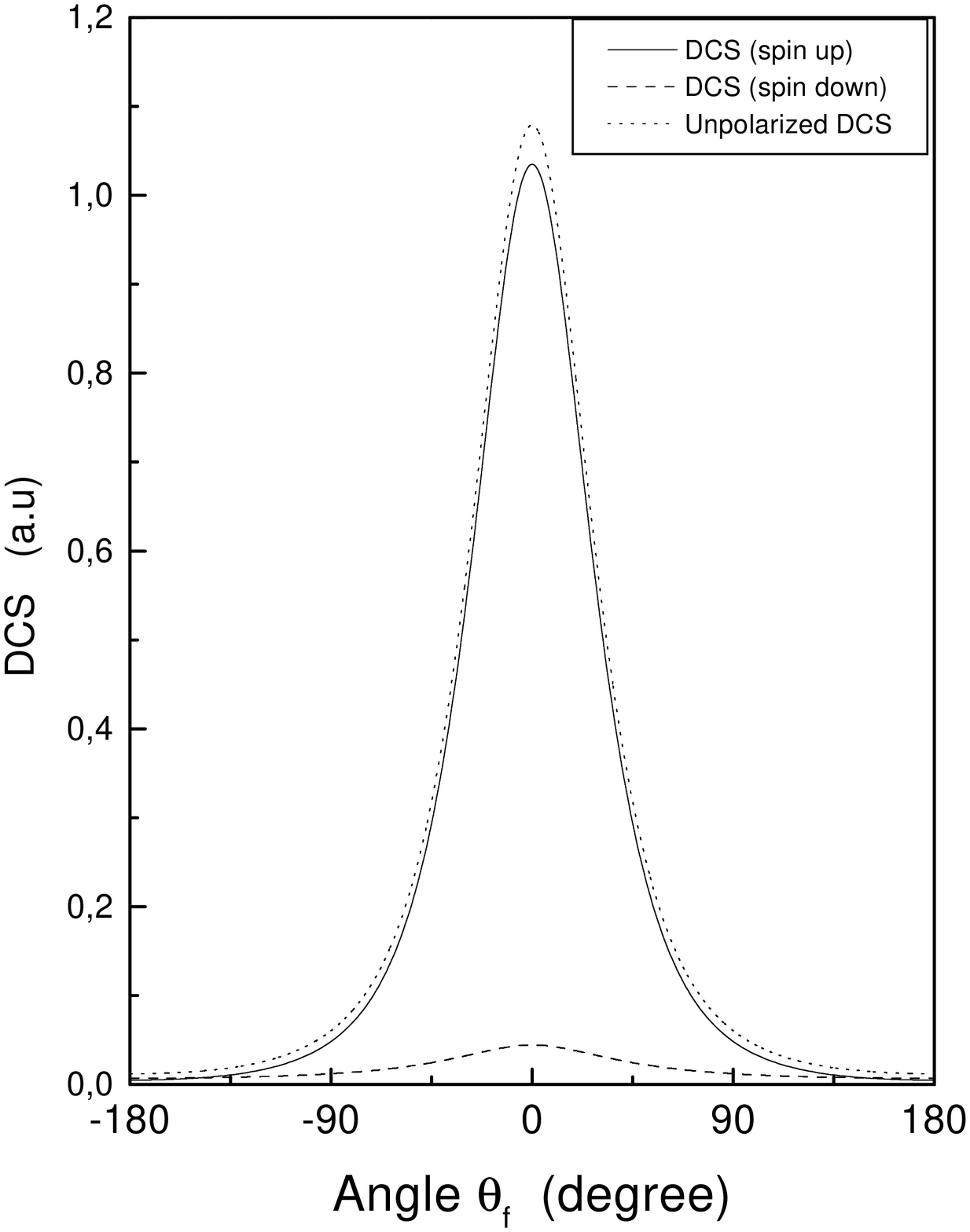,width=\linewidth,height=2.5in}
  \caption{The three DCSs : DCS $(\uparrow)$, DCS $(\downarrow)$ and unpolarized DCS scaled
   in $10^{-11}$ as functions of the angle $\theta_{f}$ in degree
   for an exchange of $\pm 100$ photons in the relativistic
    regime $\gamma=2.0$, $\textit{E}=1.0$ \label{fig8}}
 \end{minipage} \hfill
 \begin{minipage}[b]{.46\linewidth}
  \centering\epsfig{figure=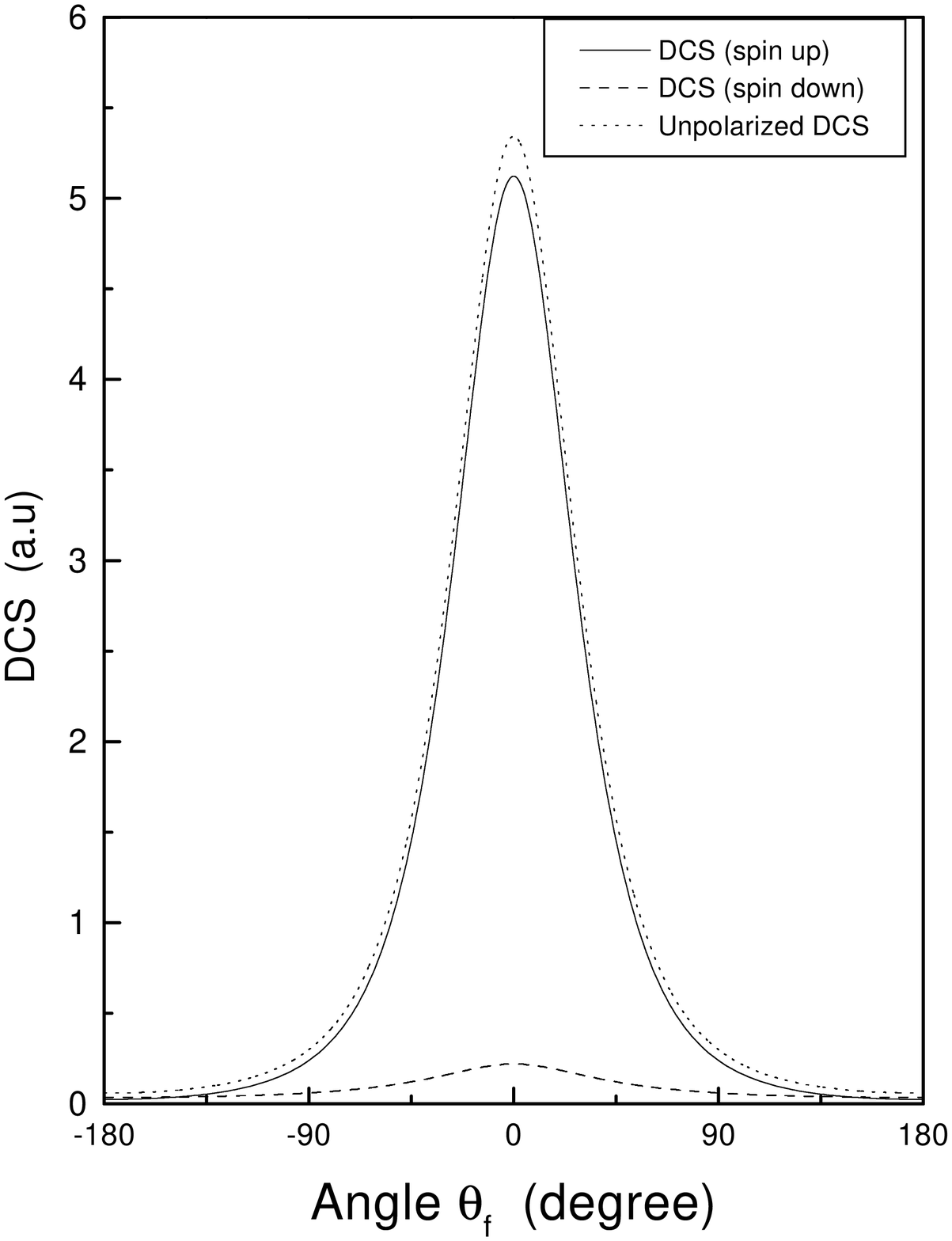,width=\linewidth,height=2.5in}
  \caption{The three DCSs : DCS $(\uparrow)$, DCS $(\downarrow)$ and unpolarized DCS scaled
   in $10^{-11}$ as functions of the angle $\theta_{f}$
   for an exchange of $\pm 500$ photons in the relativistic
    regime $\gamma=2.0$, $\textit{E}=1.0$  \label{fig9}}
 \end{minipage}
\end{figure}
 On the
other hand, there is absolutely no change in the degree of
polarization even if we sum over a much larger number of photons.
This is due to the fact that the degree of polarization $P$ can be
written as
\begin{equation}
P=1-2\frac{\frac{d\sigma (\downarrow )}{d\Omega }}{\frac{d\sigma
(\uparrow )}{d\Omega }+\frac{d\sigma (\downarrow )}{d\Omega }}
\end{equation}
and that the second term in the above equation is independent with
respect of the number of photons exchanged. Once again, one can
not distinguish between the degree of polarization corresponding
to an exchange of $\pm 100$ photons and $n=\pm 500$.
\begin{figure}[h]
 \begin{minipage}[b]{.46\linewidth}
  \centering\epsfig{figure=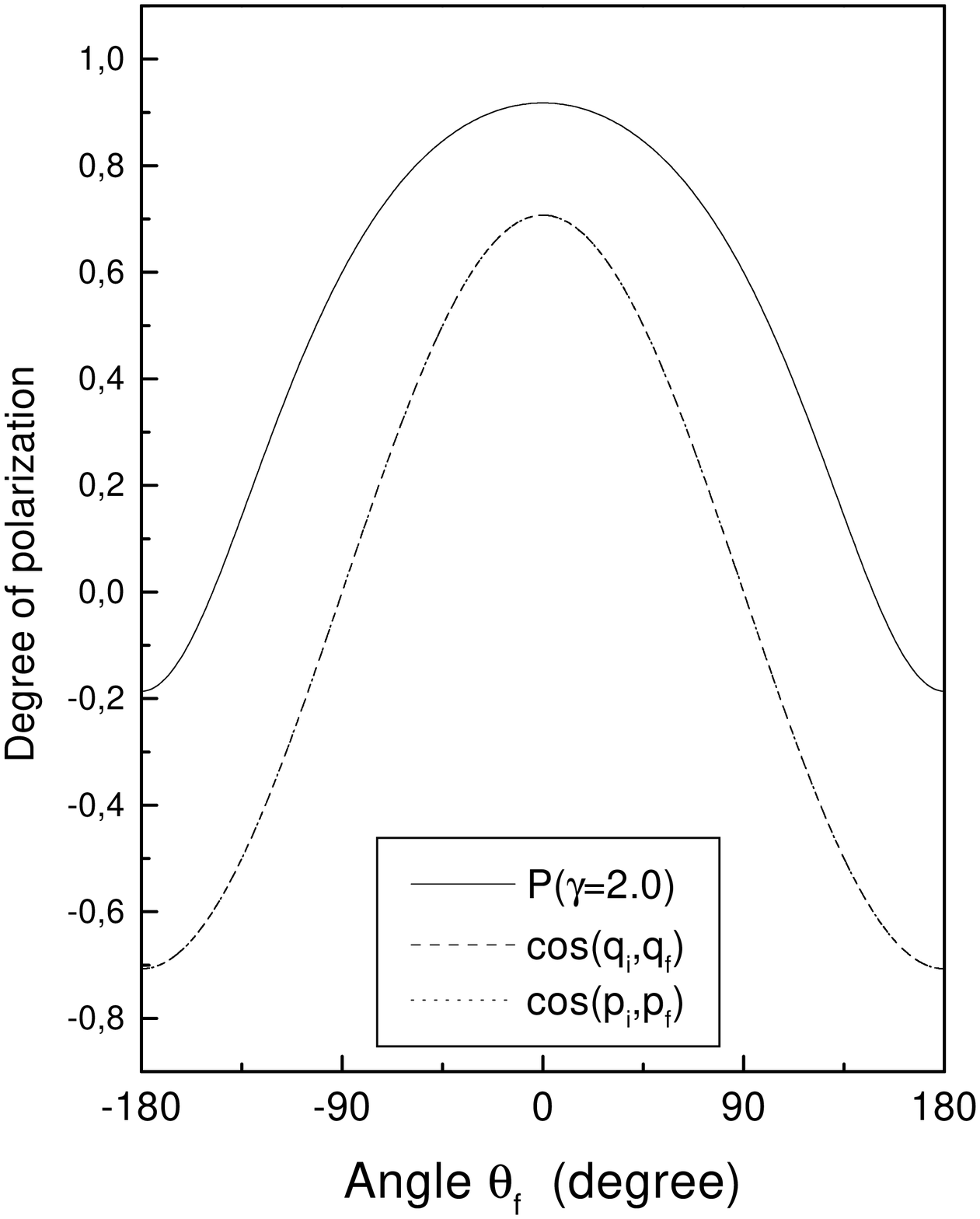,width=\linewidth,height=2.5in}
  \caption{The behaviors of the degree of polarization $P(\gamma=2.0)$,
   $cos(\mathbf{q}_i,\mathbf{q}_f)$ and $cos(\mathbf{p}_i,\mathbf{p}_f)$
    as functions of the angle $\theta_f$. \vspace{0.3cm}  \label{fig10}}
 \end{minipage} \hfill
 \begin{minipage}[b]{.46\linewidth}
  \centering\epsfig{figure=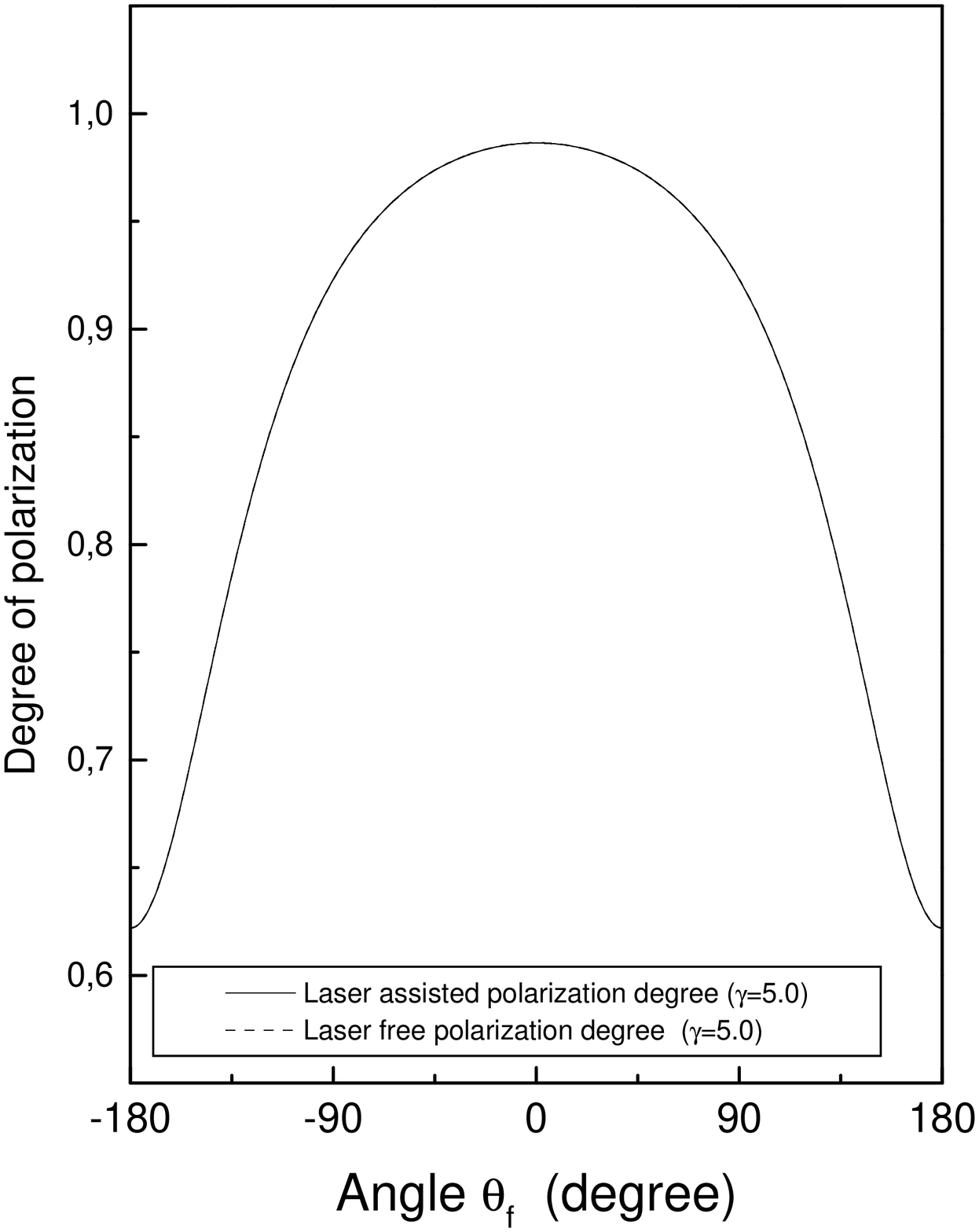,width=\linewidth,height=2.5in}
  \caption{The behaviors of the degree of polarization $P(\gamma=5.0,\quad n=\pm 1000)$
   and the laser free degree of polarization $P(\gamma=5.0)$
    as functions of the angle $\theta_f$.  \label{fig11}}
 \end{minipage}
\end{figure}
This degree of polarization is shown in Fig. 10. Finally, some
observations have to be made concerning the ultra relativistic
regime which corresponds to a relativistic parameter $\gamma =5.0$
and an electric field strength $E=5.89\,\, (a.u)$. In that case,
the cut off is very high, namely of the order of $\pm 695000$
photons exchanged and due to a lack of high speed computing
facilities, we can not sum over a very large number of photons.
However, the same conclusion holds for the degree of polarization,
which is insensitive to the number of photons exchanged. In Figure
(11), we compare the laser free degree of polarization and the
degree of polarization for an exchange of $\pm 1000$ photons. The
curves are very close. The independence of the degree of
polarization with respect to the number of photons exchanged has
been checked for various geometries leading always to the same
aforementioned conclusion except for the well known case of
elastic forward scattering where there is a divergence of the DCS.
For this particular geometry, $\theta_i=0^{\circ}$ and
$\phi_i=0^{\circ}$ while the angles of the scattered electron are
$\phi_f=180^{\circ}$ with $\theta_f$ varying from $-180^{\circ}$
to $180^{\circ}$.
\subsection{Conclusion}
In this work, we have studied the behavior of the three DCSs (the
helicity non flip polarized differential cross section, the
helicity flip polarized differential cross section and the
unpolarized differential cross section) in the absence and in
presence of a linearly polarized laser field. We have mainly
studied the non relativistic and the relativistic regime and in
all cases, the sum of the two polarized differential cross
sections always gives the unpolarized differential cross section
while the degree of polarization is independent with respect to
the number of photons exchanged and is very close to the laser
free degree of polarization. These results have been obtained in
the first order of perturbation theory and are valid for a very
wide range of angular geometries.
\begin{widetext}
\appendix
\section{Unpolarized DCS coefficients}
\begin{eqnarray}
A_1=|\mathbf{p}_f| |\mathbf{p}_i| c^{2}
\cos(\theta_{if})+c^{4}+E_f E_i
\end{eqnarray}
\begin{eqnarray}
B_1=2 (k.p_i) a^2 \left((k.p_f) c^{2}-2 E_f \omega\right)
\end{eqnarray}
\begin{equation}
C_1=2 (k.p_f) a^2 \left((k.p_i) c^{2}-2 E_i \omega\right)
\end{equation}
\begin{equation}
D_1=8 (k.p_f) (k.p_i) (a^2)^{2} \omega^{2}
\end{equation}
\begin{equation}
E_1=2 \left((a_1.p_f) (k.p_i) c^{2}-(a_1.p_i) (k.p_f) c^{2}+2
(a_1.p_i) E_f \omega\right)
\end{equation}
\begin{equation}
F_1=2 \left(-(a_1.p_f) (k.p_i) c^{2}+2 (a_1.p_f) E_i
\omega+(a_1.p_i) (k.p_f) c^{2}\right)
\end{equation}
\begin{equation}
G_1=\frac{4 a^2 \omega}{c^{2}} \left(-(k.p_f) c^{2} E_i-(k.p_i)
c^{2} E_f-|\mathbf{p}_f| |\mathbf{p}_i| c^{2} \cos(\theta_{if})
\omega-c^{4} \omega+E_f E_i \omega\right)
\end{equation}
\begin{eqnarray}
H_1&=&\frac{4}{c^{2}} \Big(2 (a_1.p_f) (a_1.p_i) c^{2}
\omega^{2}-(k.p_f) (k.p_i) a^2 c^{4}+(k.p_f) a^2 c^{2} E_i
\omega+(k.p_i) a^2 c^{2}
E_f \omega\nonumber\\
&&+a^2 |\mathbf{p}_f| |\mathbf{p}_i| c^{2} \cos(\theta_{if})
\omega^{2}+a^2 c^{4} \omega^{2}-a^2 E_f E_i \omega^{2}\Big)
\end{eqnarray}
\begin{equation}
X_1=-8 (a_1.p_f) (k.p_i) a^2 \omega^{2}
\end{equation}
\begin{equation}
Y_1=-8 (a_1.p_i) (k.p_f) a^2 \omega^{2}
\end{equation}
\section{Added polarized DCS coefficients}
\begin{eqnarray}
A_2&=&\frac{1}{c^{4}}\Big(|\mathbf{p}_f|^{2} |\mathbf{p}_i|^{2}
c^{4} \cos(\theta_{if})-|\mathbf{p}_f|^{2} c^{2} \cos(\theta_{if})
E_i^{2}+|\mathbf{p}_f| |\mathbf{p}_i| c^{6}-|\mathbf{p}_i|^{2}
c^{2} \cos(\theta_{if}) E_f^{2}\nonumber\\
&&+c^{4} \cos(\theta_{if}) E_f E_i+\cos(\theta_{if}) E_f^{2}
E_i^{2}\Big)
\end{eqnarray}
\begin{eqnarray}
B_2&=&\frac{2}{c^{6}} \Big[-2 (a_1.p_f) (a_1.p_i) |\mathbf{p}_f|
|\mathbf{p}_i| c^{4} \omega^{2}+2 (a_1.p_f) (a_1.p_i)
|\mathbf{p}_f| c^{3} \cos(\theta_i) E_i \omega^{2}-2 (a_1.p_f)\nonumber\\
&&\times (a_1.p_i) |\mathbf{p}_i| c^{3} \cos(\theta_f) E_f
\omega^{2}+2 (a_1.p_f) (a_1.p_i) c^{2} \cos(\theta_f)
\cos(\theta_i) E_f E_i
\omega^{2}+2 (a_1.p_f) \nonumber\\
&&\times(a_1.s_i) (k.p_i) |\mathbf{p}_f| c^{6} \omega+2
(a_1.p_f) (a_1.s_i) (k.p_i) c^{5} \cos(\theta_f) E_f \omega-2 (a_1.p_i) (a_1.s_f) (k.p_f)\nonumber\\
&& \times  |\mathbf{p}_i|c^{6} \omega+ 2 (a_1.p_i) (a_1.s_f)
(k.p_f) c^{5} \cos(\theta_i) E_i \omega+4 (a_1.p_i) (a_1.s_f)
|\mathbf{p}_i| c^{4} E_f \omega^{2}-4
(a_1.p_i)\nonumber\\
&& (a_1.s_f) c^{3} \cos(\theta_i) E_f E_i \omega^{2}+2 (a_1.s_f)
(a_1.s_i) (k.p_f) (k.p_i) c^{8}-4 (a_1.s_f)
(a_1.s_i) (k.p_i) c^{6} E_f \omega \nonumber\\
&&+(k.p_f) (k.p_i)a^2 |\mathbf{p}_f| |\mathbf{p}_i| c^{6}+(k.p_f)
(k.p_i) a^2 c^{4} \cos(\theta_{if}) E_f E_i-(k.p_f)
a^2 |\mathbf{p}_f| |\mathbf{p}_i| c^{4} E_i \omega\nonumber\\
&&+(k.p_f) a^2 |\mathbf{p}_f| c^{3} \cos(\theta_i) E_i^{2}
\omega-(k.p_f) a^2 |\mathbf{p}_i|^{2} c^{4} \cos(\theta_{if}) E_f
\omega+(k.p_f) a^2 |\mathbf{p}_i| c^{3} \cos(\theta_{if})\nonumber\\
&& \times\cos(\theta_i) E_f E_i \omega+(k.p_i) a^2
|\mathbf{p}_f|^{2} c^{4} \cos(\theta_{if}) E_i \omega-(k.p_i) a^2
|\mathbf{p}_f| |\mathbf{p}_i| c^{4} E_f \omega+(k.p_i)
 a^2 |\mathbf{p}_f| \nonumber\\
&&\times c^{3} \cos(\theta_{if}) \cos(\theta_f) E_f E_i
 \omega+(k.p_i) a^2 |\mathbf{p}_i| c^{3} \cos(\theta_f) E_f^{2}
 \omega-2 (k.p_i) a^2 c^{2}\cos(\theta_{if}) E_f^{2} E_i \omega\nonumber\\
&&-a^2 |\mathbf{p}_f|^{2} |\mathbf{p}_i|^{2} c^{4}
\cos(\theta_{if}) \omega^{2}+a^2 |\mathbf{p}_f|^{2} |\mathbf{p}_i|
c^{3} \cos(\theta_{if}) \cos(\theta_i) E_i
\omega^{2}-a^2 |\mathbf{p}_f| |\mathbf{p}_i|^{2} c^{3} \cos(\theta_{if})\nonumber\\
&&\times \cos(\theta_f) E_f \omega^{2}-a^2 |\mathbf{p}_f|
|\mathbf{p}_i| c^{6} \omega^{2}+ a^2 |\mathbf{p}_f| |\mathbf{p}_i|
c^{2} \cos(\theta_{if}) \cos(\theta_f) \cos(\theta_i) E_f E_i
\omega^{2}+a^2
|\mathbf{p}_f|\nonumber\\
&&\times |\mathbf{p}_i| c^{2} E_f E_i \omega^{2}+a^2
|\mathbf{p}_f| c^{5} \cos(\theta_i) E_i \omega^{2}-a^2
|\mathbf{p}_f| c \cos(\theta_i) E_f E_i^{2} \omega^{2}+2 a^2 |\mathbf{p}_i|^{2} c^{2} \cos(\theta_{if})\nonumber\\
&& \times E_f^{2} \omega^{2}-a^2 |\mathbf{p}_i| c^{5}
\cos(\theta_f) E_f \omega^{2}-2 a^2 |\mathbf{p}_i| c
\cos(\theta_{if}) \cos(\theta_i) E_f^{2} E_i \omega^{2} -a^2
|\mathbf{p}_i| c
\cos(\theta_f) E_f^{2} \nonumber\\
&&\times E_i \omega^{2}+a^2 c^{4} \cos(\theta_f) \cos(\theta_i)
E_f E_i \omega^{2}+a^2 \cos(\theta_f) \cos(\theta_i) E_f^{2}
E_i^{2} \omega^{2}\Big]
\end{eqnarray}
\begin{eqnarray}
C_2&=&\frac{2}{c^{6}} \Big[-2 (a_1.p_f) (a_1.p_i) |\mathbf{p}_f|
|\mathbf{p}_i| c^{4} \omega^{2}-2 (a_1.p_f) (a_1.p_i)
|\mathbf{p}_f| c^{3} \cos(\theta_i) E_i \omega^{2}+2 (a_1.p_f)\nonumber\\
&&\times (a_1.p_i) |\mathbf{p}_i| c^{3} \cos(\theta_f) E_f
\omega^{2}+2 (a_1.p_f) (a_1.p_i) c^{2} \cos(\theta_f)
\cos(\theta_i) E_f E_i \omega^{2}-2 (a_1.p_f)\nonumber\\
&&\times (a_1.s_i) (k.p_i) |\mathbf{p}_f| c^{6} \omega+2 (a_1.p_f)
(a_1.s_i) (k.p_i) c^{5} \cos(\theta_f) E_f \omega+4
(a_1.p_f)(a_1.s_i) |\mathbf{p}_f| \nonumber\\
&&\times c^{4} E_i \omega ^{2}-4 (a_1.p_f) (a_1.s_i) c^{3}
\cos(\theta_f) E_f E_i \omega^{2}+2
(a_1.p_i) (a_1.s_f) (k.p_f)|\mathbf{p}_i| c^{6} \omega\nonumber\\
&&+2 (a_1.p_i) (a_1.s_f) (k.p_f)
 c^{5} \cos(\theta_i) E_i \omega+2 (a_1.s_f) (a_1.s_i) (k.p_f) (k.p_i)
 c^{8}-4 (a_1.s_f)\nonumber\\
&& \times (a_1.s_i) (k.p_f) c^{6} E_i \omega+(k.p_f) (k.p_i) a^2
|\mathbf{p}_f| |\mathbf{p}_i| c^{6}+(k.p_f) (k.p_i) a^2 c^{4}
\cos(\theta_{if}) E_f E_i\nonumber\\
&&-(k.p_f) a^2 |\mathbf{p}_f| |\mathbf{p}_i| c^{4} E_i
\omega+(k.p_f) a^2 |\mathbf{p}_f| c^{3} \cos(\theta_i) E_i^{2}
\omega+(k.p_f) a^2 |\mathbf{p}_i|^{2} c^{4} \cos(\theta_{if}) E_f
\omega\nonumber\\
&&+(k.p_f) a^2 |\mathbf{p}_i| c^{3} \cos(\theta_{if})
\cos(\theta_i) E_f E_i \omega-2 (k.p_f) a^2 c^{2}
\cos(\theta_{if}) E_f E_i^{2} \omega-(k.p_i) a^2
|\mathbf{p}_f|^{2} c^{4} \nonumber\\
&&\times\cos(\theta_{if}) E_i \omega-(k.p_i) a^2 |\mathbf{p}_f|
|\mathbf{p}_i| c^{4} E_f \omega+(k.p_i) a^2 |\mathbf{p}_f| c^{3}
\cos(\theta_{if}) \cos(\theta_f) E_f E_i \omega+(k.p_i)
\nonumber\\
&&\times a^2|\mathbf{p}_i| c^{3} \cos(\theta_f) E_f^{2} \omega-a^2
|\mathbf{p}_f|^{2} |\mathbf{p}_i|^{2} c^{4} \cos(\theta_{if})
\omega^{2}-a^2 |\mathbf{p}_f|^{2} |\mathbf{p}_i| c^{3}
\cos(\theta_{if}) \cos(\theta_i) E_i \omega^{2}\nonumber\\
&&+2 a^2 |\mathbf{p}_f|^{2} c^{2} \cos(\theta_{if}) E_i^{2}
\omega^{2}+a^2 |\mathbf{p}_f| |\mathbf{p}_i|^{2} c^{3}
\cos(\theta_{if}) \cos(\theta_f) E_f \omega^{2}-a^2 |\mathbf{p}_f|
|\mathbf{p}_i|
c^{6} \omega^{2}+a^2 \nonumber\\
&& \times|\mathbf{p}_f||\mathbf{p}_i| c^{2} \cos(\theta_{if})
\cos(\theta_f) \cos(\theta_i) E_f E_i \omega ^{2}+a^2
|\mathbf{p}_f| |\mathbf{p}_i| c^{2} E_f E_i \omega^{2}-a^2
|\mathbf{p}_f| c^{5} \cos(\theta_i) E_i \nonumber\\
&&\times\omega^{2}-2 a^2 |\mathbf{p}_f| c \cos(\theta_{if})
\cos(\theta_f) E_f E_i^{2} \omega^{2}-a^2 |\mathbf{p}_f| c
\cos(\theta_i) E_f E_i^{2} \omega^{2}+a^2 |\mathbf{p}_i| c^{5}
\cos(\theta_f) E_f \omega
^{2}\nonumber\\
&&-a^2 |\mathbf{p}_i| c \cos(\theta_f) E_f^{2} E_i \omega^{2}+a^2
c^{4} \cos(\theta_f) \cos(\theta_i) E_f E_i \omega^{2}+a^2
\cos(\theta_f) \cos(\theta_i) E_f^{2} E_i^{2} \omega^{2}\Big]
\end{eqnarray}
\begin{eqnarray}
D_2&=&\frac{8 (a^2)^{2} \omega^{2}}{c^{8}} \Big(-(k.p_f) (k.p_i)
|\mathbf{p}_f| |\mathbf{p}_i| c^{6}+(k.p_f) (k.p_i) c^{4}
\cos(\theta_{if}) E_f E_i+(k.p_f) |\mathbf{p}_f|
 |\mathbf{p}_i| \nonumber\\
&&\times c^{4} E_i \omega-(k.p_f) |\mathbf{p}_f| c^{3}
 \cos(\theta_i) E_i^{2} \omega-(k.p_f) |\mathbf{p}_i|^{2} c^{4}
 \cos(\theta_{if}) E_f \omega+(k.p_f)
|\mathbf{p}_i| c^{3} \cos(\theta_{if}) \nonumber\\
&&\times \cos(\theta_i) E_f E_i \omega-(k.p_i) |\mathbf{p}_f|^{2}
c^{4} \cos(\theta_{if}) E_i \omega+(k.p_i) |\mathbf{p}_f|
|\mathbf{p}_i| c^{4} E_f \omega+(k.p_i) |\mathbf{p}_f| c^{3}
\cos(\theta_{if})\nonumber\\
&& \times\cos(\theta_f) E_f E_i \omega-(k.p_i) |\mathbf{p}_i|
c^{3} \cos(\theta_f) E_f^{2} \omega+|\mathbf{p}_f|^{2}
|\mathbf{p}_i|^{2}
 c^{4} \cos(\theta_{if}) \omega^{2}-|\mathbf{p}_f|^{2} |\mathbf{p}_i|
 c^{3} \cos(\theta_{if})  \nonumber\\
&&\times\cos(\theta_i) E_i\omega^{2}-|\mathbf{p}_f|
 |\mathbf{p}_i|^{2} c^{3} \cos(\theta_{if}) \cos(\theta_f)
 E_f \omega^{2}+|\mathbf{p}_f| |\mathbf{p}_i| c^{6} \omega^{2}
 +|\mathbf{p}_f| |\mathbf{p}_i| c^{2} \cos(\theta_{if}) \cos(\theta_f)
 \nonumber\\
&&\times \cos(\theta_i) E_f E_i
\omega^{2}-|\mathbf{p}_f||\mathbf{p}_i| c^{2} E_f E_i
\omega^{2}-|\mathbf{p}_f| c^{5} \cos(\theta_i) E_i
\omega^{2}+|\mathbf{p}_f| c \cos(\theta_i) E_f E_i^{2}
\omega^{2}-|\mathbf{p}_i|  \nonumber\\
&&\times c^{ 5} \cos(\theta_f) E_f \omega^{2}+|\mathbf{p}_i|c
\cos(\theta_f) E_f^{2} E_i \omega^{2}+c^{4} \cos(\theta_f)
\cos(\theta_i) E_f E_i \omega^{2} - \cos(\theta_f) \cos(\theta_i)
E_f^{2} \nonumber\\
&&\times E_i^{2} \omega^{2}\Big)
\end{eqnarray}
\begin{eqnarray}
E_2&=&\frac{2}{c^{ 4}} \Big((a_1.p_f) (k.p_i) |\mathbf{p}_f|
|\mathbf{p}_i| c^{4}+(a_1.p_f) (k.p_i) c^{2} \cos(\theta_{if})
E_f E_i-(a_1.p_f)|\mathbf{p}_f| |\mathbf{p}_i| c^{2} E_i \omega\nonumber\\
&&+(a_1.p_f) |\mathbf{p}_f| c \cos(\theta_i) E_i^{2}
\omega-(a_1.p_f) |\mathbf{p}_i|^{2} c^{2} \cos(\theta_{if}) E_f
\omega+(a_1.p_f)|\mathbf{p}_i| c \cos(\theta_{if})\nonumber\\
&&\times  \cos(\theta_i) E_f E_i \omega-(a_1.p_i) (k.p_f)
|\mathbf{p}_f| |\mathbf{p}_i|
c^{4}-(a_1.p_i) (k.p_f) c^{2}\cos(\theta_{if}) E_f E_i\nonumber\\
&&-(a_1.p_i) |\mathbf{p}_f|^{2} c^{2} \cos(\theta_{if}) E_i
\omega+(a_1.p_i) |\mathbf{p}_f|
|\mathbf{p}_i| c^{2} E_f \omega-(a_1.p_i) |\mathbf{p}_f| c \cos(\theta_{if})\nonumber\\
&&\times \cos(\theta_f) E_f E_i \omega-(a_1.p_i) |\mathbf{p}_i| c
\cos(\theta_f) E_f ^{2}
\omega+2 (a_1.p_i) \cos(\theta_{if}) E_f^{2} E_i \omega\nonumber\\
&&-(a_1.s_f) (k.p_i) |\mathbf{p}_f| c^{4} \cos(\theta_{if})
E_i-(a_1.s_f)
(k.p_i) |\mathbf{p}_i| c^{4} E_f +(a_1.s_f) |\mathbf{p}_f| |\mathbf{p}_i|^{2}\nonumber\\
&&\times c^{4} \cos(\theta_{if}) \omega-(a_1.s_f) |\mathbf{p}_f|
|\mathbf{p}_i|
c^{3} \cos(\theta_{if}) \cos(\theta_i) E_i \omega+(a_1.s_f) |\mathbf{p}_i| c^{6} \omega \nonumber\\
&&+(a_1.s_f) |\mathbf{p}_i| c^{2} E_f E_i \omega-(a_1.s_f) c^{5}
\cos(\theta_i) E_i \omega-(a_1.s_f) c \cos(\theta_i) E_f E_i^{2} \omega\nonumber\\
&&+(a_1.s_i) (k.p_f) |\mathbf{p}_f| c^{4} E_i+(a_1.s_i) (k.p_f)
|\mathbf{p}_i|
c^{4} \cos(\theta_{if}) E_f+(a_1.s_i) |\mathbf{p}_f|^{2} |\mathbf{p}_i| c^{4}\nonumber\\
&&\times \cos(\theta_{if}) \omega+(a_1.s_i) |\mathbf{p}_f|
|\mathbf{p}_i|
c^{3} \cos(\theta_{if}) \cos(\theta_f) E_f \omega+(a_1.s_i) |\mathbf{p}_f| c^{6} \omega\nonumber\\
&&-(a_1.s_i) |\mathbf{p}_f| c^{2} E_f E_i \omega-2 (a_1.s_i)
|\mathbf{p}_i| c^{2 }
\cos(\theta_{if}) E_f^{2} \omega+(a_1.s_i) c^{5} \cos(\theta_f) E_f \omega\nonumber\\
&&+(a_1.s_i) c \cos(\theta_f) E_f^{2} E_i \omega\Big)
\end{eqnarray}
\begin{eqnarray}
F_2&=&\frac{2}{c^{4 }} \Big(-(a_1.p_f) (k.p_i) |\mathbf{p}_f|
|\mathbf{p}_i| c^{4}-(a_1.p_f) (k.p_i) c^{2} \cos(\theta_{if})
E_f E_i+(a_1.p_f) |\mathbf{p}_f| |\mathbf{p}_i| c^{2} E_i\omega\nonumber\\
&& -(a_1.p_f) |\mathbf{p}_f| c \cos(\theta_i) E_i^{2}
\omega-(a_1.p_f) |\mathbf{p}_i|^{2} c^{2} \cos(\theta_{if}) E_f
\omega-(a_1.p_f) |\mathbf{p}_i| c \cos(\theta_{if})
\cos(\theta_i) \nonumber\\
&&\times E_f E_i \omega+2 (a_1.p_f) \cos(\theta_{if}) E_f E_i^{2}
\omega+(a_1.p_i) (k.p_f) |\mathbf{p}_f| |\mathbf{p}_i|
c^{4}+(a_1.p_i) (k.p_f)c^{2} \cos(\theta_{if}) \nonumber\\
&&\times  E_f E_i-(a_1.p_i) |\mathbf{p}_f|^{2} c^{2}
\cos(\theta_{if}) E_i \omega-(a_1.p_i) |\mathbf{p}_f|
|\mathbf{p}_i| c^{2} E_f \omega+(a_1.p_i) |\mathbf{p}_f| c
\cos(\theta_{if}) \cos(\theta_f) \omega\nonumber\\
&& \times E_f E_i+(a_1.p_i) |\mathbf{p}_i| c \cos(\theta_f)
E_f^{2} \omega+(a_1.s_f) (k.p_i) |\mathbf{p}_f|
c^{4} \cos(\theta_{if}) E_i+(a_1.s_f) (k.p_i) |\mathbf{p}_i|c^{4} E_f\nonumber\\
&&+ (a_1.s_f) |\mathbf{p}_f| |\mathbf{p}_i|^{2} c^{4}
\cos(\theta_{if}) \omega+(a_1.s_f) |\mathbf{p}_f| |\mathbf{p}_i|
c^{3} \cos(\theta_{if}) \cos(\theta_i) E_i \omega-2 (a_1.s_f)
|\mathbf{p}_f| c^{2} \cos(\theta_{if}) \nonumber\\
&&\times E_i^{2} \omega+(a_1.s_f) |\mathbf{p}_i| c^{6}
\omega-(a_1.s_f)|\mathbf{p}_i| c^{2} E_f E_i \omega+(a_1.s_f)
c^{5} \cos(\theta_i) E_i \omega +(a_1.s_f) c \cos(\theta_i) E_f
\nonumber\\
&&\times E_i^{2} \omega-(a_1.s_i) (k.p_f) |\mathbf{p}_f| c^{4}
E_i-(a_1.s_i) (k.p_f) |\mathbf{p}_i| c^{4} \cos(\theta_{if})
E_f+(a_1.s_i)
|\mathbf{p}_f| ^{2} |\mathbf{p}_i| c^{4} \cos(\theta_{if}) \omega\nonumber\\
&&-(a_1.s_i) |\mathbf{p}_f| |\mathbf{p}_i| c^{3} \cos(\theta_{if})
\cos(\theta_f) E_f \omega+(a_1.s_i) |\mathbf{p}_f| c^{6}
\omega+(a_1.s_i) |\mathbf{p}_f| c^{2} E_f
E_i \omega-(a_1.s_i) c^{5}\nonumber\\
&&\times  \cos(\theta_f) E_f \omega-(a_1.s_i) c \cos(\theta_f)
E_f^{2} E_i \omega\Big)
\end{eqnarray}
\begin{eqnarray}
G_2&=&\frac{4 a^2 \omega}{c^{6}} \Big((k.p_f) |\mathbf{p}_i|^{2}
c^{4} \cos(\theta_{if}) E_f-(k.p_f) c^{2} \cos(\theta_{if}) E_f
E_i^{2}+(k.p_i) |\mathbf{p}_f|^{2} c^{4}
\cos(\theta_{if})E_i\nonumber\\
&&-(k.p_i) c^{2} \cos(\theta_{if}) E_f^{2} E_i-|\mathbf{p}_f|^{2}
|\mathbf{p}_i|^{2} c^{4} \cos(\theta_{if})
\omega+|\mathbf{p}_f|^{2} |\mathbf{p}_i| c^{ 3} \cos(\theta_{if})
\cos(\theta_i) E_i \omega\nonumber\\
&&+|\mathbf{p}_f| |\mathbf{p}_i|^{2} c^{3} \cos(\theta_{if})
\cos(\theta_f) E_f \omega-|\mathbf{p}_f| |\mathbf{p}_i| c^{6}
\omega+|\mathbf{p}_f| c^{5} \cos(\theta_i) E_i
\omega-|\mathbf{p}_f| c \cos(\theta_{if})\nonumber\\
&& \times\cos(\theta_f) E_f E_i^{2} \omega+|\mathbf{p}_i| c^{5}
\cos(\theta_f) E_f \omega-|\mathbf{p}_i| c \cos(\theta_{if})
\cos(\theta_i) E_f^{2} E_i \omega-c^{4} \cos(\theta_{if}) E_f
\nonumber\\
&&\times E_i \omega+\cos(\theta_{if}) E_f^{2} E_i^{2} \omega\Big)
\end{eqnarray}
\begin{eqnarray}
H_2&=&\frac{4}{c^{6}} \Big(2 (a_1.p_f) (a_1.p_i) c^{2}
\cos(\theta_{if}) E_f E_i \omega^{2}-2 (a_1.p_f) (a_1.p_i) c^{2}
\cos(\theta_f) \cos(\theta_i) E_f E_i \omega^{2}-2
(a_1.p_f)\nonumber\\
&&\times (a_1.s_i) (k.p_i) c^{5} \cos(\theta_f) E_f \omega-2
(a_1.p_f) (a_1.s_i) |\mathbf{p}_i| c^{4} \cos(\theta_{if}) E_f
\omega^{2}+2 (a_1.p_f) (a_1.s_i) c^{3} \cos(\theta_f) \nonumber\\
&&\times E_f E_i \omega^{2}-2 (a_1.p_i) (a_1.s_f) (k.p_f) c^{5}
\cos(\theta_i) E_i \omega-2 (a_1.p_i) (a_1.s_f) |\mathbf{p}_f|
c^{4}
 \cos(\theta_{if}) E_i \omega^{2}+2 \nonumber\\
&&\times (a_1.p_i) (a_1.s_f) c^{3}
 \cos(\theta_i) E_f E_i \omega^{2}-2 (a_1.s_f) (a_1.s_i) (k.p_f) (k.p_i) c^{8}+2
(a_1.s_f) (a_1.s_i) (k.p_f) c^{6}\nonumber\\
&&\times E_i \omega+2 (a_1.s_f) (a_1.s_i) (k.p_i) c^{6} E_f
\omega+2 (a_1.s_f) (a_1.s_i) |\mathbf{p}_f||\mathbf{p}_i|
c^{6} \cos(\theta_{if}) \omega^{2}+2 (a_1.s_f) (a_1.s_i)\nonumber\\
&&\times c^{8} \omega^{2}-2 (a_1.s_f) (a_1.s_i) c^{4} E_f E_i
\omega^{2}-(k.p_f) (k.p_i) a^2 |\mathbf{p}_f|
|\mathbf{p}_i| c ^{6}-(k.p_f)(k.p_i) a^2 c^{4} \cos(\theta_{if}) E_f E_i\nonumber\\
&& +(k.p_f) a^2 |\mathbf{p}_f| |\mathbf{p}_i| c^{4} E_i
\omega-(k.p_f) a^2 |\mathbf{p}_f| c^{3} \cos(\theta_i) E_i^{2}
\omega-(k.p_f) a^2
|\mathbf{p}_i| c^{3} \cos(\theta_{if})\cos(\theta_i) E_f E_i \omega\nonumber\\
&&+(k.p_f) a^2 c^{2} \cos(\theta_{if}) E_f E_i ^{2} \omega+(k.p_i)
a^2 |\mathbf{p}_f| |\mathbf{p}_i| c^{4} E_f \omega-(k.p_i) a^2
|\mathbf{p}_f| c^{3} \cos(\theta_{if})
\cos(\theta_f) E_f E_i \omega\nonumber\\
&&-(k.p_i) a^2
 |\mathbf{p}_i| c^{3} \cos(\theta_f) E_f^{2}  \omega+(k.p_i)
  a^2 c^{2} \cos(\theta_{if}) E_f^{2} E_i \omega-a^2 |\mathbf{p}_f|
   |\mathbf{p}_i| c^{2} \cos(\theta_{if})
\cos(\theta_f) \cos(\theta_i) \nonumber\\
&&\times E_f E_i \omega^{2}-a^2 |\mathbf{p}_f| |\mathbf{p}_i|
c^{2} E_f E_i \omega^{2}+a^2 |\mathbf{p}_f| c \cos(\theta_{if})
\cos(\theta_f) E_f E_i^{2} \omega^{2}+a^2 |\mathbf{p}_f| c
\cos(\theta_i) E_f E_i^{2} \omega^{2} \nonumber\\
&&+a^2 |\mathbf{p}_i| c \cos(\theta_{if}) \cos(\theta_i) E_f^{2}
E_i \omega ^{2}+a^2 |\mathbf{p}_i| c \cos(\theta_f) E_f^{2} E_i
\omega^{2}+a^2 c^{4} \cos(\theta_{if}) E_f E_i \omega^{2}-a^2
c^{4}  \nonumber\\
&&\times\cos(\theta_f)\cos(\theta_i) E_f E_i \omega^{2}-a^2
\cos(\theta_{if}) E_f^{2} E_i^{2}  \omega^{2}-a^2 \cos(\theta_f)
\cos(\theta_i) E_f^{2} E_i^{2} \omega^{2}\Big)
\end{eqnarray}
\begin{eqnarray}
X_2&=&\frac{8 a^2 \omega}{c^{6}} \Big((a_1.p_f) (k.p_i)
|\mathbf{p}_i| c^{3} \cos(\theta_f) E_f \omega-(a_1.p_f) (k.p_i)
c^{2} \cos(\theta_{if}) E_f E_i \omega+ (a_1.p_f)|\mathbf{p}_i|^{2} \nonumber\\
&&\times c^{2} \cos(\theta_{if}) E_f \omega^{2}-(a_1.p_f)
|\mathbf{p}_i| c \cos(\theta_{if}) \cos(\theta_i) E_f E_i
\omega^{2}-(a_1.p_f) |\mathbf{p}_i| c \cos(\theta_f) E_f E_i
\nonumber\\
&&\times\omega^{2}+(a_1.p_f) \cos(\theta_f) \cos(\theta_i) E_f
E_i^{2} \omega^{2}+(a_1.p_i) (k.p_f) |\mathbf{p}_f| |\mathbf{p}_i|
c^{4}
\omega-(a_1.p_i) (k.p_f) |\mathbf{p}_f|  \nonumber\\
&&\times c^{3}\cos(\theta_i) E_i \omega-(a_1.p_i) |\mathbf{p}_f|
|\mathbf{p}_i| c^{2} E_f \omega^{2}+(a_1.p_i) |\mathbf{p}_f| c
\cos(\theta_i) E_f E_i \omega^{ 2}+(a_1.p_i) |\mathbf{p}_i|
 \nonumber\\
&&\times c\cos(\theta_f) E_f^{2}
\omega^{2}-(a_1.p_i)\cos(\theta_f) \cos(\theta_i) E_f^{2} E_i
\omega^{2}+(a_1.s_f) (k.p_f) (k.p_i)
|\mathbf{p}_i| c^{6}-(a_1.s_f)  \nonumber\\
&&\times (k.p_f) |\mathbf{p}_i| c^{4} E_i \omega+(a_1.s_f) (k.p_f)
c^{3} \cos(\theta_i) E_i^{2} \omega+(a_1.s_f)(k.p_i)
|\mathbf{p}_f| c^{4} \cos(\theta_{if}) E_i \omega\nonumber\\
&&-(a_1.s_f) (k.p_i) |\mathbf{p}_i| c^{4} E_f \omega-(a_1.s_f)
|\mathbf{p}_f| |\mathbf{p}_i|^{2} c^{4} \cos(\theta_{if})
\omega^{2}
+(a_1.s_f)|\mathbf{p}_f||\mathbf{p}_i| c^{3} \cos(\theta_{if})\nonumber\\
&& \times \cos(\theta_i) E_i \omega^{2}-(a_1.s_f)|\mathbf{p}_i|
c^{6} \omega^{2}+(a_1.s_f)
|\mathbf{p}_i| c^{2} E_f E_i \omega^{2}+(a_1.s_f) c^{5} \cos(\theta_i) E_i \omega^{2}\nonumber\\
&&-(a_1.s_f) c \cos(\theta_i) E_f E_i^{2} \omega^{2}-(a_1.s_i)
(k.p_f) (k.p_i) |\mathbf{p}_f| c^{6}+(a_1.s_i) (k.p_i)
 |\mathbf{p}_f| c^{4} E_f \omega\nonumber\\
&&-(a_1.s_i) (k.p_i) c^{3} \cos(\theta_f) E_f^{2} \omega\Big)
\end{eqnarray}
\begin{eqnarray}
Y_2&=&\frac{8 a^2 \omega}{c^{6}} \Big((a_1.p_f) (k.p_i)
|\mathbf{p}_f| |\mathbf{p}_i| c^{4} \omega-(a_1.p_f) (k.p_i)
|\mathbf{p}_i| c^{3} \cos(\theta_f) E_f \omega-(a_1.p_f)
|\mathbf{p}_f|
 |\mathbf{p}_i| c^{2} E_i \omega^{2}\nonumber\\
&&+(a_1.p_f) |\mathbf{p}_f|
 c \cos(\theta_i) E_i^{2} \omega^{2}+(a_1.p_f) |\mathbf{p}_i|
 c \cos(\theta_f) E_f E_i \omega^{2}-
(a_1.p_f) \cos(\theta_f) \cos(\theta_i) E_f E_i^{2}
\omega^{2} \nonumber\\
&&+(a_1.p_i) (k.p_f) |\mathbf{p}_f|c^{3} \cos(\theta_i) E_i
\omega-(a_1.p_i) (k.p_f) c^{2} \cos(\theta_{if}) E_f E_i
\omega+(a_1.p_i) |\mathbf{p}_f|^{2} c^{2} \cos(\theta_{if})\nonumber\\
&&\times E_i \omega^{2}-(a_1.p_i) |\mathbf{p}_f| c
\cos(\theta_{if}) \cos(\theta_f) E_f E_i \omega ^{2}-(a_1.p_i)
|\mathbf{p}_f| c \cos(\theta_i) E_f E_i \omega^{2}+(a_1.p_i)\nonumber\\
&&\times\cos(\theta_f) \cos(\theta_i) E_f^{2} E_i
\omega^{2}-(a_1.s_f) (k.p_f) (k.p_i) |\mathbf{p}_i|
c^{6}+(a_1.s_f) (k.p_f)
|\mathbf{p}_i| c^{4} E_i \omega-(a_1.s_f) \nonumber\\
&&\times(k.p_f) c^{3} \cos(\theta_i) E_i^{2} \omega+(a_1.s_i)
(k.p_f) (k.p_i) |\mathbf{p}_f| c^{6}-(a_1.s_i) (k.p_f)
|\mathbf{p}_f| c^{4} E_i \omega+(a_1.s_i)\nonumber\\
&&\times (k.p_f) |\mathbf{p}_i| c^{4} \cos(\theta_{if}) E_f
\omega-(a_1.s_i) (k.p_i) |\mathbf{p}_f| c^{4} E_f \omega+(a_1.s_i)
(k.p_i) c^{3} \cos(\theta_f) E_f^{2}
\omega \nonumber\\
&&-(a_1.s_i) |\mathbf{p}_f|^{2}|\mathbf{p}_i| c^{4}
\cos(\theta_{if}) \omega^{2}+(a_1.s_i) |\mathbf{p}_f|
|\mathbf{p}_i| c^{3 } \cos(\theta_{if}) \cos(\theta_f) E_f
\omega^{2}-(a_1.s_i) |\mathbf{p}_f|\nonumber\\
&&\times c^{6} \omega^{2}+(a_1.s_i) |\mathbf{p}_f| c^{2} E_f E_i
\omega^{2}+(a_1.s_i) c^{5} \cos(\theta_f) E_f \omega^{2}-(a_1.s_i)
c \cos(\theta_f) E_f^{2} E_i \omega^{2}\Big)
\end{eqnarray}
\end{widetext}

\end{document}